\documentclass[12pt]{article}

\usepackage[centertags]{amsmath}
\usepackage{amsfonts}
\usepackage{amssymb}
\usepackage{amsthm}
\usepackage{amsmath}
\usepackage{graphicx,color}
\usepackage{ulem}
\usepackage{upgreek}

\newcommand{\beq}{\begin{equation}}
\newcommand{\eeq}[1]{\label{#1}\end{equation}}
\newcommand{\bea}{\begin{eqnarray}}
\newcommand{\eea}[1]{\label{#1}\end{eqnarray}}
\newcommand{\sea}[2]{\begin{subequations}\label{#1}\begin{align} #2 \end{align}\end{subequations}}
\newcommand{\ba}{\begin{align}}
\newcommand{\ea}{\end{align}}

\newcommand{\marg}[1]{}%\marginpar{\textbf{\textcolor{red}{#1}}}

\newcommand{\hc}{\textrm{h.c.}}
\newcommand{\rf}[1]{(\ref{#1})}

\renewcommand{\overleftarrow}[1]{\overset\leftarrow #1} % find something more alike to \vec{•}
\def\tf{\tfrac}
\def\de{\partial}
\def\ds{\displaystyle{\not{\!\partial\!\,}}}
\def\pss{\displaystyle{\not{\!\psi\!\,}}}
\def\Rs{\displaystyle{\not{\;\!\!\!R}}}
\def\Rs{\displaystyle{\not{\;\!\!\!R}}}
\def\Gs{\displaystyle{\not{\;\!\!\!G}}}
\def\gt{\tilde{g}}
\def\a{\alpha}
\def\b{\beta}
\def\g{\gamma}
\def\G{\Gamma}
\def\d{\delta}
\def\D{\Delta}

\def\ve{\varepsilon}

\def\h{\eta}

\def\l{\lambda}
\def\L{\Lambda}
\def\m{\mu}
\def\n{\nu}

\def\r{\rho}

\def\s{\sigma}

\def\t{\tau}

\def\vf{\varphi}
\def\ps{\psi}
\def\Ps{\Psi}

\begin{document}
\setlength{\topmargin}{-1cm} \setlength{\oddsidemargin}{0cm}
\setlength{\evensidemargin}{0cm}
\begin{titlepage}
\begin{center}
{\Large \bf Gravitational Interactions of Higher-Spin Fermions}

\vspace{25pt}

{\large Marc \textsc{Henneaux}, Gustavo \textsc{Lucena G\'omez} and Rakibur \textsc{Rahman}}

\vspace{25pt}

Physique Th\'eorique et Math\'ematique \& International Solvay Institutes\\
Universit\'e Libre de Bruxelles, Campus Plaine C.P. 231, B-1050 Bruxelles, Belgium

\end{center}
\vspace{20pt}

\begin{abstract}

We investigate the cubic interactions of a massless higher-spin fermion with gravity in flat space and present
covariant $2-s-s$ vertices, compatible with the gauge symmetries of the system, preserving parity. This explicit
construction relies on the BRST deformation scheme that assumes locality and Poincar\'e invariance. Consistent
nontrivial cubic deformations exclude minimal gravitational coupling and may appear only with a number of
derivatives constrained in a given range. Derived in an independent manner, our results do agree with those obtained
from the light-cone formulation or inspired by string theory. We also show that none of the Abelian vertices deform
the gauge transformations, while all the non-Abelian ones are obstructed in a local theory beyond the cubic order.

\end{abstract}

\end{titlepage}

\newpage

%%%%%%%%%%%%%%%%%%%%%%%%%%%%%%%%%%%%%%%
%%%%%%%%%%%%%%%%%%%%%%%%%%%%%%%%%%%%%%%
\section{Introduction}\label{sec:intro}
%%%%%%%%%%%%%%%%%%%%%%%%%%%%%%%%%%%%%%%
%%%%%%%%%%%%%%%%%%%%%%%%%%%%%%%%%%%%%%%

Interactions of massless higher-spin fields in flat space are severely constrained. Powerful no-go theorems forbid
minimal coupling to gravity in Minkowski space when the particle's spin exceeds the value
$s=2$~\cite{No-go,Aragone,Porrati}. Higher-spin particles may still possess gravitational multipoles. For bosonic
fields, for example, consistent $2-s-s$ trilinear vertices exist~\cite{2-3-3,2-s-s}, albeit they may get obstructed
beyond the cubic order in a local theory.

In general, one can consider $s_1-s_2-s_3$ cubic vertices involving massless fields of arbitrary spins. The number
of derivatives in these vertices, as the light-cone formulation elucidates, is restricted, which provides a way of
classifying them~\cite{Metsaev_Boson,Metsaev}. The complete list of such vertices for bosonic fields appeared already
in~\cite{Cubic-general}, but their explicit construction may call for the Noether procedure~\cite{Karapet} or
the BRST-BV cohomological methods~\cite{2-3-3,2-s-s,BRST-BV}. The tensionless limit of open string
theory, on the other hand, yields a set of cubic vertices in one-to-one correspondence with that found in the
light-cone formulation~\cite{Taronna}.

This paper is devoted to the study of the cubic interactions of an arbitrary-spin massless fermion with dynamical
gravity in flat space of dimension $D\geq4$, and is a sequel to its electromagnetic counterpart~\cite{EMours}.
Fermions are interesting in that they are required by supersymmetry$-$a crucial ingredient of string theory that
indeed incorporates an infinite tower of higher-spin fields. Their appearance in the higher-spin
literature~\cite{Metsaev,Taronna,EMours,FV} is yet meager; our study aims at filling some of the gaps.
For totally symmetric Dirac fermions $\ps_{\m_1...\m_n}$, of spin $s=n+\tf{1}{2}$, we employ the powerful BRST
deformation scheme~\cite{BRST-BV} for systematic construction of covariant interaction vertices allowed by
gauge symmetry. The underlying assumptions include only locality, Poincar\'e invariance and conservation of
parity. The covariant $2-s-s$ cubic vertices we are about to find will thus complement their bosonic counterparts
constructed in Ref.~\cite{2-s-s}. We will consider only $s\geq\tf{5}{2}$, since spin $\tf{3}{2}$ has no consistency
issues with gravity (gauge deformations of the free system indeed leads uniquely to $\mathcal N=1$
supergravity~\cite{N=1SUGRA} under reasonable assumptions).

The organization of the paper is as follows: In the remaining of this Section we clarify our conventions
and notations, and spell out our main results. Section~\ref{sec:BRST} is a brief account of the BRST deformation
scheme~\cite{BRST-BV} for irreducible gauge theories$-$a machinery to be used in the remaining of the paper. In
Section~\ref{sec:5/2}, we consider in great detail $s=\tf{5}{2}$, which serves as a prototype for arbitrary spin.
Treating the gauge-algebra-deforming/preserving cases separately, we explicitly construct all the $2-\tf{5}{2}-\tf{5}{2}$
vertices, and cast them into various off-shell forms to make some desired properties manifest. Section~\ref{sec:arbitrary}
is a straightforward arbitrary-spin generalization that mimics the spin-$\tf{5}{2}$ case. In Section~\ref{sec:2ndorder}
we show that our non-Abelian vertices face obstructions in a local theory beyond the cubic order. We conclude with
some remarks in Section~\ref{sec:remarks}. Three appendices supplement the main text to provide useful
technical details much required throughout the bulk of the paper.

%%%%%%%%%%%%%%%%%%%%%%%%%%%%%%%%%%%%%%%%%%%%%%%%%%%%%%%%%%%%
\subsection*{Conventions \& Notations}\label{subsec:convnot}
%%%%%%%%%%%%%%%%%%%%%%%%%%%%%%%%%%%%%%%%%%%%%%%%%%%%%%%%%%%%

We work with mostly positive metric in Minkowski spacetime of dimension $D\geq4$. The Clifford algebra is
$\{\g^\m,\g^\n\} \equiv +2\h^{\m\n}$, and the $\g$-matrices obey $\g^{\m\,\dagger} \equiv \h^{\m\m}\g^\m$.
The Dirac adjoint is defined as $\bar{\ps}_\m=\ps^\dagger_\m\g^0$. The Levi-Civita tensor,
$\epsilon_{\m_1\m_2...\m_D}$, is normalized as $\epsilon_{01...(D-1)}\equiv +1$. Totally antisymmetric
product of $\g$-matrices have unit weight: $\g^{\m_1....\m_n}\equiv\g^{[\m_1}\g^{\m_2}...\g^{\m_n]}$,
with $[i_1...i_n]$ denoting a totally antisymmetric expression in all the indices $i_1,...,i_n$ with a
normalization factor $\tf{1}{n!}$. The totally symmetric expression $(i_1...i_n)$ comes with the same
normalization. The anticommutator of two antisymmetric products of $\g$-matrices is denoted as follows:
$\g^{\m_1...\m_m,\,\n_1...\n_n}\equiv\tf{1}{2}\{\g^{\m_1...\m_m},\g^{\n_1...\n_n}\}$. We will use the
symbol $\h^{\m\n|\r\s}\equiv\h^{[\m\n]|[\r\s]}=\h^{\r\s|\m\n}$ to denote the tensor
$\tf{1}{2}\left(\h^{\m\r}\h^{\n\s}-\h^{\m\s}\h^{\n\r}\right)$.

The spin-2 graviton field will be denoted by $h_{\m\n}$, while its 1-curl by a Fraktur letter:
$\mathfrak{h}_{\m\n\Vert\r}\equiv 2\de_{[\m}h_{\n]\r}$. The  2-curl of the graviton is simply the
linearized Riemann tensor, denoted by $R_{\m\n\r\s}$ as usual:
$R_{\m\n}{}^{\r\s}\equiv 4\de_{[\m}\de^{[\r}h_{\n]}{}^{\s]}$. Its trace is the linearized Ricci
tensor: $R_{\m\n}\equiv R_{\m\r\n}{}^\r$, whose trace in turn is the Ricci scalar: $R\equiv R^\m_\m$.
The symbol $\Rs_{\m\n}$ will denote the double $\g$-trace, $\g^{\r\s}R_{\m\n\r\s}$, of the Riemann
tensor.\footnote{In all other cases ``slash'' will always mean a single $\g$-trace:
$\g^\m Q_\m\equiv\displaystyle{\not{\!Q}}$, and ``prime'' a trace.} We will also use the symbols:
$R^{+\m\n\a\b}\equiv\left(\h^{\m\n|\r\s}+\tf{1}{2}\g^{\m\n\r\s}\right)R_{\r\s}{}^{\a\b}$ and
$\mathfrak h^{+\m\n\Vert\l}\equiv\left(\h^{\m\n|\r\s}+\tf{1}{2}\g^{\m\n\r\s}\right)\mathfrak h_{\r\s\Vert}{}^{\l}$.

For the spin-$\tf{5}{2}$ field, $\ps_{\m\n}$, we denote a 1-curl by an upright greek letter:
$\uppsi_{\m\n\Vert\r}\equiv 2\de_{[\m}\ps_{\n]\r}$, and a 2-curl (curvature tensor) by an uppercase greek
letter: $\Ps_{\m\n|}{}^{\r\s}\equiv 4\de_{[\m}\de^{[\r}\ps_{\n]}{}^{\s]}$.

For arbitrary spin $s=n+\tf{1}{2}$, we have a totally symmetric rank-$n$ tensor-spinor $\ps_{\n_1...\n_n}$,
whose curvature is a rank-$2n$ tensor-spinor, $\Ps_{\m_1\n_1|\m_2\n_2|...|\m_n\n_n}$, defined as the $n$-curl,
\beq \Ps_{\m_1\n_1|\m_2\n_2|...|\m_n\n_n}\equiv\left[...\left[\,\left[\de_{\m_1}...\de_{\m_n}\ps_{\n_1...\n_n}
-(\m_1\leftrightarrow\n_1)\right]-(\m_2\leftrightarrow\n_2)\right]...\right]-(\m_n\leftrightarrow\n_n).
\nonumber\eeq{null1}
This is the Weinberg curvature tensor~\cite{Curvature}. Its properties and relation to the equations of motion
(EoMs), along with those of the Riemann tensor, will be discussed in Appendix~\ref{sec:curvatures}.
However, for $s>\tf{5}{2}$, one can have multiple intermediate curls. An $m$-curl for $0<m<n$, will be
denoted by an upright greek letter with an explicit superscript $m$,\footnote{One may extrapolate $m$ to include
the values $n$ and $0$: $m=n$ gives nothing but the curvature tensor, $\uppsi^{(n)}_{\m_1\n_1|...|\m_n\n_n}
=\Ps_{\m_1\n_1|...|\m_n\n_n}$, whereas $m=0$ corresponds to the original field itself, $\uppsi^{(0)}_{\n_1...\n_n}
=\ps_{\n_1...\n_n}$.}
\beq \uppsi^{(m)}_{\m_1\n_1|...|\m_m\n_m\Vert\n_{m+1}...\n_n}\equiv\left[...\left[\,\left[\de
_{\m_1}...\de_{\m_m}\ps_{\n_1...\n_n}-(\m_1\leftrightarrow\n_1)\right]-(\m_2\leftrightarrow\n_2)
\right]...\right]-(\m_m\leftrightarrow\n_m).\nonumber\eeq{null2}
The Fronsdal tensor for the fermionic field~\cite{Fronsdal} will be denoted by $\mathcal S_{\m_1...\m_n}$, i.e.,
\beq \mathcal S_{\m_1...\m_n}=i\left[\ds\,\ps_{\m_1...\m_n}-n\de_{(\m_1}\pss_{\m_2...\m_n)}\right].\nonumber
\eeq{null3}
Finally, the symbol ``$\doteq$'' will mean the equality of expressions up to a total derivative, while
``$\approx$'' the equivalence of vertices up to field redefinitions and total derivatives.

%%%%%%%%%%%%%%%%%%%%%%%%%%%%%%%%%%%%%%%%%%%
\subsection*{Results}\label{subsec:results}
%%%%%%%%%%%%%%%%%%%%%%%%%%%%%%%%%%%%%%%%%%%

\begin{itemize}
 \item We provide a cohomological proof of the well-known fact that in flat space a massless spin-$\tf{5}{2}$ field cannot
 have minimal coupling to gravity~\cite{Aragone,Porrati}. This result generalizes easily to higher-spin fermions.
 \item For spin $s=n+\tf{1}{2}$, we find that the possible number of derivatives in a cubic $2-s-s$ vertex is restricted to
 only five allowed values: $2n-2, 2n-1, 2n, 2n+1$ and $2n+2$, with only one inequivalent vertex for each value. Derived
 independently, this is in complete accordance with the light-cone-formulation results of Metsaev~\cite{Metsaev}.
 \item Two of the vertices$-$those with the lowest $2n-2$ and $2n-1$ number of derivatives$-$are non-Abelian, while the other
 three are Abelian.
 \item Only two of these vertices exist in $D=4$: the ones with the lowest $2n-2$ and highest $2n+2$ number of derivatives.
 All five vertices are nontrivial in $D\geq5$.
 \item None of the Abelian vertices deform the gauge transformations. The highest-derivative one can be written in
 a strictly gauge-invariant 3-curvature term, while the other two can be rendered gauge invariant only up to total derivatives.
 \item In a local theory, with no additional dynamical degrees of freedom, the non-Abelian vertices get obstructed beyond
 the cubic order.
\end{itemize}

A summary of our results, for the prototypical example of spin $\tf{5}{2}$, appears below:

\begin{table}[ht]
\caption{Prototypical Example of $2-\tf{5}{2}-\tf{5}{2}$ Vertices with $p$ Derivatives}
\vspace{6pt}
\centering
\begin{tabular}{c c c c}
\hline\hline\vspace{4pt}
~$p$&~~~~Vertex~~~~~&~~Abelian?~~&Exists in\\

\hline\vspace*{-10pt}\\
2 & $i\bar\ps_{\m\a}R^{+\m\n\a\b}\ps_{\n\b}+\tf{i}{2}\bar{\pss}_\m\Rs^{\m\n}\pss_\n+\tf{i}{4}h_{\m\n}
\bar\uppsi_{\r\s\Vert\l}\,\g^{\m\r\s\a\b,\,\n\l\g}\,\uppsi_{\a\b\Vert\g}$ & No & $D\geq4$\\\\
3 & $i\bar{\uppsi}_{\m\n\Vert}{}^\r\left(\mathfrak{h}^+_{\r\s\Vert\l}\,\g^{\l\m\n\a\b}+\g^{\l\m\n\a\b}\,
\mathfrak{h}^+_{\r\s\Vert\l}\right)\uppsi_{\a\b\Vert}{}^\s$ & No & $D\geq5$\\\\
4 & $ih_{\m\n}\bar\Ps_{\r\s|\t\l}\,\g^{\m\r\s\a\b,\,\n\t\g}\,\Ps_{\a\b|\g}{}^\l$ & Yes & $D\geq5$\\\\
5 & $i\mathfrak{h}_{\m\n\Vert\l}\bar\Ps^{\m\t|}{}_{\r\s}\,\g^{\l\r\s\a\b}\,\Ps^\n{}_{\t|\a\b}$ & Yes & $D\geq5$\\\\
\vspace{2pt}
6 & $iR_{\m\n\r\s}\bar\Ps^{\r\s|\a\b} \Ps_{\a\b}{}^{\m\n}$ & Yes & $D\geq4$\\
\hline\hline\vspace{6pt}
\end{tabular}
\end{table}

%%%%%%%%%%%%%%%%%%%%%%%%%%%%%%%%%%%%%%%%%%%%%%%%%%%%%
%%%%%%%%%%%%%%%%%%%%%%%%%%%%%%%%%%%%%%%%%%%%%%%%%%%%%
\section{The BRST Deformation Scheme}\label{sec:BRST}
%%%%%%%%%%%%%%%%%%%%%%%%%%%%%%%%%%%%%%%%%%%%%%%%%%%%%
%%%%%%%%%%%%%%%%%%%%%%%%%%%%%%%%%%%%%%%%%%%%%%%%%%%%%

In this Section we outline the BRST deformation scheme$-$our tool to find gravitational vertices of massless
higher-spin fermions.\footnote{This account is an almost verbatim repetition of that appearing in~\cite{EMours},
where electromagnetic couplings of massless higher-spin fermions were considered.} As pointed out in~\cite{BRST-BV},
one can reformulate the classical problem of introducing consistent interactions in a gauge theory in terms of the
BRST differential and the BRST cohomology. The advantage is that the search for all possible consistent interactions
becomes systematic, thanks to the cohomological approach. Obstructions to deforming a gauge-invariant action also
become related to precise cohomological classes of the BRST differential.

%%%%%%%%%%%%%%%%%%%%%%%%%%%%%%%%%%%%%%
\subsubsection*{Fields and Antifields}
%%%%%%%%%%%%%%%%%%%%%%%%%%%%%%%%%%%%%%

Let us consider an irreducible gauge theory of a collection of fields $\{\phi^i\}$, with $m$ gauge invariances,
$\d_{\ve}\phi^i=R^i_\a\ve^\a, \a=1,2,...,m$. Corresponding to each gauge parameter
$\ve^\a$, one introduces a ghost field $\mathcal C^\a$, with the same algebraic symmetries but
opposite Grassmann parity ($\epsilon$). The original fields and ghosts are collectively called fields, denoted
by $\Phi^A$. The configuration space is further enlarged by introducing, for each field and ghost, an antifield
$\Phi^*_A$, that has the same algebraic symmetries (in its indices when $A$ is a multi-index) but opposite
Grassmann parity.

%%%%%%%%%%%%%%%%%%%%%%%%%
\subsubsection*{Gradings}
%%%%%%%%%%%%%%%%%%%%%%%%%

In the algebra generated by the fields and antifields, we introduce
two gradings: the pure ghost number ($pgh$) and the antighost number ($agh$). The former is non-zero only for the
ghost fields. In particular, for irreducible gauge theories, $pgh(\mathcal C^\a)=1$, while $pgh(\phi^i)=0$ for
any original field. The antighost number, on the other hand, is non-zero only for the antifields $\Phi^*_A$.
Explicitly, $agh(\Phi^*_A)=pgh(\Phi^A)+1,~agh(\Phi^A)=0=pgh(\Phi^*_A)$. The ghost number ($gh$) is another grading,
defined as $gh=pgh-agh$.

%%%%%%%%%%%%%%%%%%%%%%%%%%%%
\subsubsection*{Antibracket}
%%%%%%%%%%%%%%%%%%%%%%%%%%%%

One defines an odd symplectic structure$-$the antibracket$-$on the space of fields and antifields:
\beq (X,Y)\equiv\frac{\d^RX}{\d\Phi^A}\frac{\d^LY}{\d\Phi^*_A}-\frac{\d^RX}{\d\Phi^*_A}\frac{\d^LY}{\d\Phi^A}\,.
\eeq{antibracket}
This definition gives $\left(\Phi^A,\Phi^*_B\right)=\d^A_B$, which is real. Because a field and its antifield have
opposite Grassmann parity, it follows that if $\Phi^A$ is real, $\Phi^*_B$ must be purely imaginary, and vice versa.
Note that the antibracket satisfies the graded Jacobi identity.

%%%%%%%%%%%%%%%%%%%%%%%%%%%%%%
\subsubsection*{Master Action}
%%%%%%%%%%%%%%%%%%%%%%%%%%%%%%

The original gauge-invariant action $S^{(0)}[\phi^i]$ is then extended to a new action $S[\Phi^A,\Phi^*_A]$, called
the master action, that includes terms involving ghosts and antifields,
\beq S[\Phi^A,\Phi^*_A]=S^{(0)}[\phi^i]+\phi^*_iR^i_\a\mathcal C^\a+\dots\,,\eeq{S_0}
which, by virtue of the Noether identities and the higher-order gauge-structure equations, satisfies the classical
master equation \beq (S,S)=0.\eeq{master} In other words, the master action $S$ incorporates compactly all the
consistency conditions pertaining to the gauge transformations.

%%%%%%%%%%%%%%%%%%%%%%%%%%%%%%%%%%
\subsubsection*{BRST Differential}
%%%%%%%%%%%%%%%%%%%%%%%%%%%%%%%%%%

The master action also plays role as the generator of the BRST differential $\mathfrak s$, which is defined as
\beq \mathfrak{s}X\equiv(S,X).\eeq{brst1}
Notice that $S$ is BRST-closed, as a simple consequence of the master equation. From the properties of the
antibracket, it also follows that $\mathfrak s$ is nilpotent, \beq \mathfrak s^2=0.\eeq{brst2} Therefore, the
master action $S$ belongs to the cohomology of $\mathfrak s$, denoted as $H(\mathfrak s)$, in the space of
local functionals of the fields, antifields, and their finite number of derivatives.

%%%%%%%%%%%%%%%%%%%%%%%%%%%%%%%%%%%%%%%
\subsubsection*{Deformed Master Action}
%%%%%%%%%%%%%%%%%%%%%%%%%%%%%%%%%%%%%%%

As we know, the existence of the master action $S$ as a solution of the master equation is completely equivalent
to the gauge invariance of the original action $S^{(0)}[\phi^i]$. Therefore, one can reformulate the problem of
introducing consistent interactions in a gauge theory as that of deforming the solution $S$ of the master equation.
Let $S$ be the solution of the deformed master equation, $(S,S)=0$. This must be a deformation of the
solution $S_0$ of the master equation of the free gauge theory, in the deformation parameter $g$,
\beq S=S_0+gS_1+g^2S_2+\mathcal O(g^3).\eeq{brst5} The master equation for $S$ splits, up to $\mathcal O(g^2)$,
into \bea (S_0,S_0)&=&0,\label{brst6.1}\\(S_0,S_1)&=&0,\label{brst6.2}\\(S_1,S_1)&=&-2(S_0,S_2).\eea{brst6.3}
Eq.~(\ref{brst6.1}) is fulfilled by assumption, and in fact $S_0$ is the generator of the BRST differential
for the free theory, which we will denote as $s$. Thus, Eq.~(\ref{brst6.2}) translates to
\beq s S_1=0,\eeq{brst1st} i.e., $S_1$ is BRST-closed.

%%%%%%%%%%%%%%%%%%%%%%%%%%%%%%%%%%%%%%%%%
\subsubsection*{First-Order Deformations}
%%%%%%%%%%%%%%%%%%%%%%%%%%%%%%%%%%%%%%%%%

If the first-order local deformations are given by
$S_1=\int a$, where $a$ is a top-form of ghost number 0, then one has the cocycle condition \beq sa\doteq0.
\eeq{cocycle0} Non-trivial deformations therefore belong to $H^0(s|d)$$-$the cohomology of the zeroth-order
BRST differential $s$, modulo total derivatives $d(\dots)$, at ghost number 0. Now, if one makes an antighost-number
expansion of the local form $a$, it stops at $agh=2$~\cite{2-s-s,BBH,gravitons}, \beq a=a_0+a_1+a_2, \qquad
agh(a_i)=i=pgh(a_i).\eeq{brst7} For cubic deformations $S_1=\int a$, it is easy to check that indeed one cannot construct
an object with $agh>2$~\cite{2-s-s}. The result, however, is more general and holds in fact also for higher order
deformations, as it follows from the results of Refs.~\cite{BBH,gravitons,N=1SUGRA}.

The significance of the various terms is worth recalling: $a_0$ is the deformation of the Lagrangian, while $a_1$
and $a_2$ encode information about the deformations of the gauge transformations and the gauge algebra
respectively~\cite{BRST-BV}. Thus, if $a_2$ is not trivial, the algebra of the gauge transformations is deformed and
becomes non-Abelian. On the other hand, if $a_2 = 0$ (up to redefinitions), the algebra remains Abelian to first order
in the deformation parameter. In that case, if $a_1$ is not trivial, the gauge transformations are deformed (remaining
Abelian), while if $a_1 = 0$ (up to redefinitions), the gauge transformations remain the same as in the undeformed case.

%%%%%%%%%%%%%%%%%%%%%%%%%%%%%%%%%%%%
\subsubsection*{Consistency Cascade}
%%%%%%%%%%%%%%%%%%%%%%%%%%%%%%%%%%%%

The various gradings are of relevance as $s$ decomposes into the sum of the Koszul-Tate differential,
$\D$, and the longitudinal derivative along the gauge orbits, $\G$: \beq s=\D+\G. \eeq{brst3}
The operator $\D$ implements the EoMs by acting only on the antifields. It decreases the antighost
number by one unit while keeping unchanged the pure ghost number. $\G$ acts only on the original fields and
produces the gauge transformations. It increases the pure ghost number by one unit without modifying the antighost
number. Accordingly, all three $\D$, $\G$ and $s$ increase the ghost number by one unit,
$gh(\D)=gh(\G)=gh(s)=1$. Note that $\D$ and $\G$ are nilpotent and anticommuting,
\beq \G^2=\D^2=0,\qquad\G\D+\D\G=0.\eeq{brst4}

Given the expansion~(\ref{brst7}) and the decomposition~(\ref{brst3}), the cocycle condition~(\ref{cocycle0}) yields
the following cascade of relations, that a consistent deformation must obey:
\bea \G a_2&\doteq&0,\label{cocycle1}\\
\D a_2+\G a_1&\doteq&0,\label{cocycle2}\\
\D a_1+\G a_0&\doteq&0.\eea{cocycle3}
We will call the set of conditions~\rf{cocycle1}--\rf{cocycle3}
the consistency cascade. Note that $a_2$ can always be chosen as $\G\text{-closed}$, instead of $\G\text{-closed}$
modulo $d$~\cite{BBH}.

%%%%%%%%%%%%%%%%%%%%%%%%%%%%%%%%%%%%%%%%%
\subsubsection*{Second-Order Deformations}
%%%%%%%%%%%%%%%%%%%%%%%%%%%%%%%%%%%%%%%%%

Finally, while the graded Jacobi identity for the antibracket renders $(S_1,S_1)$ BRST-closed, the second-order
consistency condition~\rf{brst6.3} requires that it actually be $s$-exact: \beq (S_1,S_1)=-2s S_2.\eeq{brst2nd}
This condition determines whether or not, in a local theory, a consistent first-order deformation gets obstructed
at the second order. Such higher-order obstructions are controlled by the local BRST cohomology group
$H^1(s|d)$~\cite{BBH}.

%%%%%%%%%%%%%%%%%%%%%%%%%%%%%%%%%%%%%%%%%%%%%%%
\subsubsection*{Non-Triviality of Deformations}
%%%%%%%%%%%%%%%%%%%%%%%%%%%%%%%%%%%%%%%%%%%%%%%

The highest-order term $a_2$ will be trivial (removable by redefinitions) iff one can get rid of it by adding to $a$
an $s$-exact term modulo $d$, $sm+dn$. Expanding $m$ and $n$ according to the antighost number, and taking into account
the fact that $m$ and $n$ also stop at $agh=2$ since they are both cubic, one finds that $a_2$ is trivial iff
$a_2=\G m_2 + dn_2$. We see that the cohomology of $\G$ modulo $d$ plays an important role. The cubic vertex will deform
the gauge algebra if and only if $a_2$ is a non-trivial element of the cohomology of $\G$ modulo $d$.

If $a_2$ is trivial, the vertex is called Abelian. In this case, one can always choose $a_2=0$, and $\G a_1=0$~\cite{BBH}.
The vertex deforms the gauge transformations unless $a_1$ is $\D$-exact modulo $d$, $a_1=\D m_2+dn_1$, where $m_2$ can be
assumed to be invariant~\cite{BBH,gravitons,N=1SUGRA}. In that instance, one can remove $a_1$, and so one can take $a_0$
to be $\G$-closed modulo $d$: the vertex only deforms the action without deforming the gauge transformations.

Non-trivial Lagrangian deformations $a_0$ are non-trivial elements in $H(\D)$, whereas trivial interactions are given by
$\D$-exact terms modulo total derivatives. Therefore, two vertices are equivalent iff they differ by $\D$-exact
terms up to total derivatives.

%%%%%%%%%%%%%%%%%%%%%%%%%%%%%%%%%%%%%%%%%%%%%%%%%%%%%%%%%%%%%%%%%%%%
%%%%%%%%%%%%%%%%%%%%%%%%%%%%%%%%%%%%%%%%%%%%%%%%%%%%%%%%%%%%%%%%%%%%
\section{Gravitational Coupling of Massless Spin 5/2}\label{sec:5/2}
%%%%%%%%%%%%%%%%%%%%%%%%%%%%%%%%%%%%%%%%%%%%%%%%%%%%%%%%%%%%%%%%%%%%
%%%%%%%%%%%%%%%%%%%%%%%%%%%%%%%%%%%%%%%%%%%%%%%%%%%%%%%%%%%%%%%%%%%%

In this Section we construct parity-preserving off-shell $2-\tf{5}{2}-\tf{5}{2}$ vertices by employing the BRST-BV
cohomological methods. The spin-$\tf{5}{2}$ system is simple enough so that one can implement the BRST-deformation
scheme with ease, while it captures many non-trivial features that will propagate along as one moves on to arbitrary
spin, which will be considered in the next Section.

The starting point is the free theory, which contains a graviton field $h_{\m\n}$ and a massless spin-$\tf{5}{2}$
tensor-spinor field $\ps_{\m\n}$, described by the action
\beq S^{(0)}[h_{\m\n},\ps_{\m\n}]=\int d^Dx\left[G^{\m\n}h_{\m\n}+\tf{1}{2}\left(\bar{\mathcal R}^{\m\n}\ps_{\m\n}
-\bar{\ps}_{\m\n}\mathcal{R}^{\m\n}\right)\right],\eeq{freeaction5/2}
which enjoys two Abelian gauge invariances \beq \d_\l h_{\m\n}=2\de_{(\m}\l_{\n)},\qquad\d_\ve\ps_{\m\n}=2\de_{(\m}\ve_{\n)},
\quad\text{with}\;\displaystyle{\not{\!\ve}}=0.\eeq{gaugeinv5/2}

For the Grassmann-even bosonic gauge parameter $\l_\m$, we introduce the Grassmann-odd bosonic ghost $C_\m$. Corresponding
to the Grassmann-odd fermionic gauge parameter $\ve_\m$, we have the Grassmann-even fermionic ghost $\xi_\m$, which is of
course $\g$-traceless. Therefore, the set of fields becomes
\beq \Phi^A=\{h_{\m\n}, C_\m, \ps_{\m\n}, \xi_\m\}.\eeq{fieldsset5/2}
For each of these fields, we introduce an antifield with the same algebraic symmetries in its indices but opposite Grassmann
parity, the set of which is
\beq \Phi^*_{A}=\{h^{*\m\n}, C^{*\m}, \bar{\ps}^{*\m\n},\bar{\xi}^{*\m}\}.\eeq{antifieldsset5/2}

Now we construct the free master action $S_0$, which is an extension of the original gauge-invariant action~\rf{freeaction5/2}
by terms involving ghosts and antifields. Explicitly,
\beq S_0=\int d^Dx\left[G^{\m\n}h_{\m\n}+\tf{1}{2}\left(\bar{\mathcal R}^{\m\n}\ps_{\m\n}-\bar{\ps}_{\m\n}\mathcal{R}^{\m\n}
\right)-2h^{*\m\n}\de_\m C_\n +(\bar{\ps}^{*\m\n}\de_\m\xi_\n-\de_\m\bar{\xi}_\n\ps^{*\m\n})\right],\eeq{freemasteraction5/2}
which is easily seen to satisfy the master equation $(S_0,S_0)=0$. Notice that the antifields appear as sources
for the ``gauge'' variations, with gauge parameters replaced by corresponding ghosts. We spell out in Table 2 the different
gradings and Grassmann parity of the various fields and antifields, along with the action of $\G$ and $\D$ on them.
\begin{table}[ht]
\caption{Properties of the Various Fields \& Antifields ($n=2$)}
\vspace{6pt}
\centering
\begin{tabular}{c c c c c c c}
\hline\hline
$Z$ &$\G(Z)$~~~&~~~$\D(Z)$~~~&$pgh(Z)$ &$agh(Z)$ &$gh(Z)$ &$\epsilon(Z)$\\ [0.5ex]
\hline
$h_{\m\n}$ & $2\de_{(\m} C_{\n)}$ & 0 & 0 & 0 & 0 & 0\\
$C_\m$ & 0 & 0 & 1 & 0 & 1 & 1\\
$h^{*\m\n}$ & 0 & $G^{\m\n}$ & 0 & 1 & $-1$ & 1\\
$C^{*\m}$ & 0 & $-2\de_\n h^{*\m\n}$ & 0 & 2 & $-2$ & 0\\ \hline
$\ps_{\m\n}$ & $2\de_{(\m}\xi_{\n)}$ & 0 & 0 & 0 & 0 & 1\\
$\xi_\m$ & 0 & 0 & 1 & 0 & 1 & 0\\
$\bar{\ps}^{*\m\n}$ & 0 & $\bar{\mathcal R}^{\m\n}$ & 0 & 1 & $-1$ & 0\\
$\bar{\xi}^{*\m}$ & 0 & $2\de_\n\bar{\chi}^{*\m\n}$ & 0 & 2 & $-2$ & 1\\
\hline\hline
\end{tabular}
\end{table}
\vspace{6pt}

For the spin-$\tf{5}{2}$ field, the Fronsdal tensor is $\mathcal S_{\m\n}=i\left[\ds\,\ps_{\m\n}-2\de_{(\m}\pss_{\n)}\right]$,
and it is related to the original EoMs via
\beq \mathcal{R}^{\m\n}=\mathcal{S}^{\m\n}-\g^{(\m}\displaystyle{\not{\!\mathcal{S}}}^{\n)}-\tf{1}{2}\h^{\m\n}\mathcal{S}',
\qquad \mathcal{S}'\equiv\mathcal{S}^\m_\m.\eeq{EoM5/2}
Note that the divergence $\de_\n\mathcal R^{\m\n}$ is not zero$-$unlike that of the Einstein tensor$-$but is
proportional to $\gamma^\m$.\footnote{The action is still gauge invariant, thanks to the $\g$-tracelessness of the gauge
parameter $\ve_\m$.} Because of this, when $\D$ acts on the fermionic antighost $\bar\xi^{*\m}$, the result is more than
a simple divergence of the antifield $\bar\ps^{*\m\n}$ (see Appendix~\ref{sec:curvatures}). Explicitly,
\beq \D\bar\xi^{*\m}=2\de_\n\bar{\chi}^{*\m\n},\qquad \bar{\chi}^{*\m\n}\equiv\bar{\ps}^{*\m\n}-\tfrac{1}{D}
\displaystyle{\not{\!\bar\ps}}^{*\n}\g^\m.\eeq{deltaofxistar}

The cohomology of $\G$ is isomorphic to the space of functions of (see Appendix~\ref{sec:cohomology})
\begin{itemize}
 \item The undifferentiated ghosts $\{C_\m, \xi_\m\}$. Also the 1-curl of the bosonic ghost $\mathfrak C_{\m\n}$
 as well as the $\g$-traceless part of the 1-curl of the fermionic ghost $\upxi_{\m\n}$,
 \item The antifields $\{h^{*\m\n}, C^{*\m}, \bar{\ps}^{*\m\n},\bar{\xi}^{*\m}\}$ and their derivatives,
 \item The curvatures $\{R_{\m\n\r\l}, \Ps_{\m\n|\r\l}\}$ and their derivatives,
 \item The Fronsdal tensor $\mathcal S_{\m\n}$ and its symmetrized derivatives.
\end{itemize}

%%%%%%%%%%%%%%%%%%%%%%%%%%%%%%%%%%%%%%%%%%%%%%%%%%%%%%%%%
\subsection{Non-Abelian Vertices}\label{subsec:nonAbelian}
%%%%%%%%%%%%%%%%%%%%%%%%%%%%%%%%%%%%%%%%%%%%%%%%%%%%%%%%%

Non-Abelian vertices are those that deform the gauge algebra. They correspond to deformations of the master action with
nontrivial terms at $agh=2$. In other words, $a_2$ is a nontrivial element in $H(\G|d)$. Notice that $a_2$ is Grassmann
even, hermitian and has $gh(a_2)=0$. Besides, we require that $a_2$ be a parity-even Lorentz scalar.

It is then clear that any $a_2$ will consist of a single antighost and two ghost fields. Let us note that two $a_2$'s are
equivalent iff they differ by $\G$-exact terms modulo total derivatives. Without loss of generality, we can thus choose
the antighost to be undifferentiated. Furthermore, any derivative acting on the ghost fields $\{C_\m,\xi_\m\}$ can be
realized as a 1-curl $\{\mathfrak C_{\m\n},\upxi_{\m\n}\}$ up to irrelevant $\G$-exact terms
(see Appendix~\ref{sec:cohomology}). Because the derivative of a ghost-curl is $\G$-exact, a nontrivial $a_2$ can
never contain more than 2 derivatives. This already poses an upper bound of 3 on the number of derivatives in a
non-Abelian vertex.

To be more explicit, let us write down all the inequivalent $a_2$'s. In view of the actions of $\G$ and $\D$ on various
(anti)fields, given any $a_2$, the consistency cascade~\rf{cocycle1}--\rf{cocycle3} unambiguously counts the number of
derivatives $p$ contained in the corresponding vertex $a_0$. Thus we can classify $a_2$'s based on the value of $p$.
The set of all possible nontrivial $a_2$'s falls into two subsets: Subset-1 contains the bosonic antighost $C^*_\m$,
while subset-2 contains the fermionic one $\xi^*_\m$. In subset-1 we have
\beq a_2=\begin{cases}p=1:\quad ig\,C^{*\m}\bar\xi^\a\g_\m\xi_\a\\
p=2:\quad ig\,C^{*\m}\bar\upxi_{\m\n}\xi^\n+\hc\\
p=3:\quad ig\,C^{*\m}\bar\upxi^{\a\b}\g_\m\upxi_{\a\b}.\end{cases}\eeq{5halfa2-C}
It is easy to see that this list is indeed complete. First, it follows from Lorentz invariance that if $p$ is odd(even),
the number of $\g$ matrices is also odd(even). The latter can be chosen simply to be 1(0). This is because if more $\g$
matrices are there, one can anti-commute them past each other using the Clifford algebra to see that only terms with 1(0)
$\g$-matrix survive, while other terms are either killed ($\displaystyle\not{\!\xi\!}=0$) or made trivial
($\g^\a\upxi_{\a\b}=\G\text{-exact}$).

Note that the $p=1$ candidate, $igC^{*\m}\bar\xi^\a\g_\m\xi_\a$, is easily ruled out as inconsistent. To see this,
we simply take its $\D$ variation and integrate by parts to find
$\D a_2\doteq 2igh^{*\m\n}\de_\n\left(\bar\xi^\a\g_\m\xi_\a\right)$, which contains nontrivial elements of $H(\G|d)$,
involving the ghost-curl $\bar\upxi_{\a\n}$. Therefore, the consistency condition~\rf{cocycle2} cannot be satisfied.

Next we consider subset-2 whose $a_2$'s contain the (undifferentiated) fermionic antighost. Again, the $a_2$'s can be
classified based on the value $p$ of the number of derivatives in the corresponding vertex $a_0$. The complete list is
\beq a_2=\begin{cases}p=0:\quad g\,\bar\xi^{*\m}\g^\a\xi_\m C_\a+\hc\\
p=1:\quad g\,\bar\xi^{*\m}\left(\xi^\n\mathfrak C_{\m\n}+\a_1\upxi_{\m\n}C^\n+\a_2\g^{\a\b}\xi_\m\mathfrak C_{\a\b}\right)+\hc\\
p=2:\quad g\,\bar\xi^{*\m}\g^\a\upxi^\b{}_\m\mathfrak{C}_{\a\b}+\hc\,,\end{cases}\eeq{5halfa2}
where $\a_1$ and $\a_2$ are dimensionless constants. Because both $\displaystyle\not{\!\xi\!}$ and $\displaystyle\not{\!\xi^*\!}$
vanish, and $\g^\a\upxi_{\a\b}=\G\text{-exact}$, any $\g$-matrix must be contracted with the bosonic ghost or with its curl.
Then one can easily verify that the list~\rf{5halfa2} indeed gives all possible inequivalent Lorentz scalars.

Here it is easy to rule out the $p=0$ candidate, $g\,\bar\xi^{*\m}\g^\a\xi_\m C_\a+\hc$, as inconsistent. Again, we simply take
its $\D$ variation and integrate by parts to obtain $\D a_2\doteq -2g\,\bar\chi^{*\m\n}\de_\n\left(\g^\a\xi_\m C_\a\right)+\hc$,
which contains nontrivial elements of $H(\G|d)$, involving the ghost-curls $\upxi_{\n\m}$ and $C_{\n\a}$. The consistency
condition~\rf{cocycle2} cannot then be satisfied.

%%%%%%%%%%%%%%%%%%%%%%%%%%%%%%%%%%%%%%%%%%%%%%%%%%%%%%%%%%%%%%%%%%%%%%%%%%%%%
\subsubsection{Minimal Coupling and Absence Thereof}\label{subsubsec:minimal}
%%%%%%%%%%%%%%%%%%%%%%%%%%%%%%%%%%%%%%%%%%%%%%%%%%%%%%%%%%%%%%%%%%%%%%%%%%%%%

A possible minimal coupling would correspond to a 1-derivative vertex. The most general $a_2$ can be written as
(dropping the already-ruled-out candidate containing $C^*_\m$)
\beq a_2=g\,\bar\xi^{*\m}\left(\xi^\n\mathfrak C_{\m\n}+\a_1\upxi_{\m\n}C^\n+\a_2\g^{\r\s}\xi_\m
\mathfrak C_{\r\s}\right)+\hc\,,\eeq{minimal-a2}
where $\a_1$ and $\a_2$ are dimensionless constants. Then we have
\beq \D a_2\doteq\G\text{-exact}-g\,\bar\chi^{*\m\a}\left(\upxi_\a{}^\n\mathfrak C_{\m\n}+\a_1\upxi_{\m\n}
\mathfrak C_\a{}^\n+\a_2\g^{\r\s}\upxi_{\a\m}\mathfrak C_{\r\s}\right)+\hc\,,\eeq{minimal-a21}
where we recall that $\bar{\chi}^{*\m\a}\equiv\bar{\ps}^{*\m\a}-\tfrac{1}{D}\displaystyle{\not{\!\bar\ps}}^{*\a}\g^\m$.
The nontrivial elements of $H(\G|d)$ appearing on the right-hand side can actually be canceled
by the choice $\a_1=-1$ and $\a_2=\tf{1}{4}$. The only subtlety are the terms containing the $\g$-trace
$\displaystyle{\not{\!\bar\ps}}^{*\a}$ of the fermionic antifield, for which one needs to use the identity:
$\g^\m\g^{\r\s}=\g^{\r\s}\g^\m+4\h^{\m[\r}\g^{\s]}$.
With the cocycle condition~\rf{cocycle2} thus satisfied, the unambiguous piece in $a_1$ reads
\beq \hat a_1=-2\left(g\,\bar\chi^{*\m\r}\uppsi_{\m\n\Vert\r}C^\n+\hc\right)+\mathcal{Y}^{\m\n}\mathfrak C_{\m\n}
+\dots\,,\eeq{a_11der}
where the ellipses stand for terms with the fermionic ghost $\xi_\m$ but not $C_\m$. This gives
\beq \hat\b^\m\equiv\frac{\d}{\d C_\m}\D\hat a_1=\left(2g\D\bar\chi^*_{\a\b}\,\uppsi^{\m\a\Vert\b}+\hc\right)
+2\D\de_\n\mathcal{Y}^{[\m\n]}.\eeq{testII1}
Similarly, because the ambiguity $\tilde a_1$ belongs to $H(\G)$, we have
\beq \tilde\b^\m\equiv\frac{\d}{\d C_\m}\D\tilde a_1=\G\text{-closed}.\eeq{testII2}
Now the cocycle condition~\rf{cocycle3} is fulfilled if
\beq \D\hat a_1+\D\tilde a_1\doteq-\G a_0\doteq 2C_\m\de_\n T^{\m\n}+\dots\,,\eeq{testII30}
for some $a_0\doteq h_{\m\n}T^{\m\n}$. Taking a functional derivative w.r.t. $C_\m$ then yields
\beq \hat\b^\m+\tilde\b^\m=2\de_\n T^{\m\n}.\eeq{testII3}
Using Eqs.~\rf{testII1} and~\rf{testII2}, and taking a $\G$ variation one is lead to the necessary condition
\beq \G\hat\b^\m=\de^\b\left[2g\,\D\bar\chi^*_{\a\b}\,\upxi^{\a\m}+\hc\right]+\de_\n\left(2\G\D\mathcal{Y}^{[\m\n]}\right)
=\de_\n\left(2\G T^{\m\n}\right).\eeq{testII4}
In $D\geq4$, this condition can never be satisfied, since the terms inside the brackets are not $\G$-exact modulo $d$. Thus we 
conclude that there is no $1$-derivative $\tf{5}{2}-\tf{5}{2}-2$ vertex; i.e., a massless spin-$\tf{5}{2}$ field cannot have 
minimal coupling to gravity in flat space~\cite{Aragone}.

%%%%%%%%%%%%%%%%%%%%%%%%%%%%%%%%%%%%%%%%%%%%%%%%%%%%%%%%%%%%%
%%%%%%%%%%%%%%%%%%%%%%%%%%%%%%%%%%%%%%%%%%%%%%%%%%%%%%%%%%%%%
\subsubsection{The 2-Derivative Vertex}\label{subsubsec:2der}
%%%%%%%%%%%%%%%%%%%%%%%%%%%%%%%%%%%%%%%%%%%%%%%%%%%%%%%%%%%%%
%%%%%%%%%%%%%%%%%%%%%%%%%%%%%%%%%%%%%%%%%%%%%%%%%%%%%%%%%%%%%

Having ruled out minimal coupling, we are lead to consider the next possibility$-$the 2-derivative vertex, for which
the corresponding $a_2$ reads
\beq a_2=\left[ig\,C^{*\m}\bar\upxi_{\m\n}\xi^\n+\hc\right]+\left[\tilde{g}\,\mathfrak{C}_{\m\n}
\bar\xi^*_\r\g^{\m\n\r\a\b}\upxi_{\a\b}+\hc\right],\eeq{1dera2}
where the coupling constants $g$ and $\tilde g$ are a priori complex, but will soon be required to be real.
Notice that, for future convenience, we wrote the term with fermionic antighost with five $\g$-matrices, instead of just
one, as it appears in Eq.~\rf{5halfa2}. The equivalence of the two forms, although rather obvious, is made explicit
in Appendix~\ref{subsec:C2d} for interested readers.
To find a possible $a_1$, we take the $\D$ variation of Eq.~\rf{1dera2} and integrate by parts:
\beq \D a_2\doteq 2\left[ig\,h^{*\m\n}\de_\n\bar\upxi_{\m\l}\xi^\l+\hc\right]+2\left[\tilde{g}
\,\bar\chi^*_{\r\s}\de^\s\left(\mathfrak{C}_{\m\n}\g^{\m\n\r\a\b}\upxi_{\a\b}\right)+\hc\right].\eeq{Delta1dera2}
In view of Eqs.~\rf{ghcurl} and~\rf{gxicurl}, the $\G$-exactness of the second piece on the right-hand side is manifest,
while in the first piece one can also use Eq.~\rf{g2} to extract $\G$-exact terms. The contributions that are nontrivial
in $H(\G)$ cancel each other only if $g$ is real. Therefore, the cocycle condition~\rf{cocycle2} is satisfied. This
gives, up to an ambiguity $\tilde a_1$,
\beq a_1=a_{1g}+a_{1\tilde g}+\tilde a_1,\eeq{2da1}
where $\G\tilde a_1=0$, and the other terms are unambiguously determined to be
\bea a_{1g}&=&ig\,h^{*\m\n}\left(\bar\upxi_{\m\l}\ps_\n{}^\l+\bar\ps_\n{}^\l\upxi_{\m\l}-2\bar\xi^\l\uppsi_{\m\l\Vert\n}
-2\bar\uppsi_{\m\l\Vert\n}\xi^\l\right),\label{a1g}\\a_{1\tilde g}&=&2\tilde{g}\left(\mathfrak{C}_{\m\n}\bar\chi^*_{\r\s}
\g^{\m\n\r\a\b}\uppsi_{\a\b\Vert}{}^\s-\mathfrak{h}_{\m\n\Vert}{}^\s\bar\chi^*_{\r\s}\g^{\m\n\r\a\b}\upxi_{\a\b}\right)
+\hc\,.\eea{a1gt}
We will now compute the $\D$ variations of the above quantities. From Eq.~\rf{a1g} one finds
\beq \D a_{1g}\doteq ig\,\bar\xi_\l\left[2G^{\m\n}\de^\l\ps_{\m\n}-3\de_\m\left(G^{\m\n}\ps_\n{}^\l\right)+\de^\n
\left(G^{\l\m}\ps_{\m\n}\right)\right]+\hc\,,\eeq{Da1g} which does not contain the bosonic ghost $C_\m$. Note
that neither can $\D\tilde a_1$ give rise to terms containing $C_\m$. This is because, if the ambiguity $\tilde a_1$
contains $C_\m$ or its curl, it must also contain the Fronsdal tensors\footnote{It cannot contain only curvatures,
because then there are too many derivatives in $\D\tilde a_1$ to possibly correspond to a vertex with $p=2$.} and
thus be $\D$-exact, so that $\D\tilde a_1=0$. This fact puts restrictions on $\D a_{1\tilde g}$\,: it may contain $C_\m$
only in the form of symmetrized derivatives, $\de_{(\m}C_{\n)}$, up to total-derivative terms. Otherwise, $\D a_1$ will
have nontrivial pieces belonging to $H(\G|d)$, and the condition $\D a_1\doteq-\G a_0$ may never be satisfied.

With the above facts in mind, we compute the following quantity, that will be useful:
\beq \b^\m_C~\equiv~\frac{\d}{\d C_\m}\D a_{1\tilde g}~=~-4\gt\D\de_{[\n}\bar\chi^*_{\r]\s}\g^{\m\n\r\a\b}
\uppsi_{\a\b\Vert}{}^\s~-~4\gt^*\bar\uppsi_{\a\b\Vert}{}^\s\g^{\m\n\r\a\b}\D\de_{[\n}\chi^*_{\r]\s}.\eeq{beta1}
The right-hand side, if non-zero, must be the divergence of a symmetric tensor: $\de_\n\mathcal X^{\m\n}$
with $\mathcal X^{\m\n}=\mathcal X^{\n\m}$. As shown in Appendix~\ref{subsec:C2d}, this is possible only if $\gt$
is real, and it yields
\beq \mathcal X^{\m\n}=2i\gt\,\bar\uppsi_{\r\s\Vert\l}\,\g^{\m\r\s\a\b,\,\n\l\g}\,\uppsi_{\a\b\Vert\g}
+(\m\leftrightarrow\n).\eeq{beta5}
Then, the bosonic ghost $C_\m$ will appear in $\D a_{1\tilde g}$ only through $\G$-exact pieces. Explicitly,
\beq \D a_{1\gt}+\G\left(\tf{1}{2}h_{\m\n}\mathcal{X}^{\m\n}\right)\doteq \tf{1}{2}h_{\m\n}\G\mathcal{X}^{\m\n}
-2\tilde{g}\left(\mathfrak{h}_{\m\n\Vert}{}^\s\D\bar\chi^*_{\r\s}\g^{\m\n\r\a\b}\upxi_{\a\b}+\hc\right).\eeq{beta6}
One can now simplify the right-hand side, which does not contain the bosonic ghost $C_\m$, but just the fermionic
one $\xi_\m$. The result is (see Appendix~\ref{subsec:C2d})
\beq \D a_{1\gt}+\G\left(\tf{1}{2}h_{\m\n}\mathcal{X}^{\m\n}\right)\doteq-i\gt\,\bar\xi_\l\left(R_{\m\n\r\s}
\g^{\m\n\l\a\b,\,\t\r\s}\,\uppsi_{\a\b\Vert\t}\right)+\hc\,.\eeq{beta6.5}
It is easy to see that the right-hand side is a nontrivial element of $H(\G|d)$. Only if it can be written, up to
$\G$-exact pieces and total derivatives, in terms of $\D\tilde a_1$ plus possibly $\D a_{1g}$, for some choice of
$\gt$, can one fulfill the condition $\D a_1\doteq-\G a_0$ and thus obtain a vertex. After a tedious but
straightforward calculation, shown in Appendix~\ref{subsec:C2d}, one can write
\beq \D a_{1\gt}+\G\left(\tf{1}{2}h_{\m\n}\mathcal{X}^{\m\n}\right)\doteq-8i\gt\,\G\left(\bar\ps_{\m\a}R^{+\m\n\a\b}
\ps_{\n\b}+\tf{1}{2}\bar{\pss}_\m\Rs^{\m\n}\pss_\n\right)-\D\mathfrak{a},\eeq{beta7}
where $R^{+\m\n\a\b}\equiv R^{\m\n\a\b}+\tf{1}{2}\g^{\m\n\r\s}R_{\r\s}{}^{\a\b}$, and $\D\mathfrak{a}$ is given in
Eq.~\rf{Dfraka}. The next step is to relate the latter quantity with $\D\tilde a_1$ and $\D a_{1g}$ up to total derivatives.
Indeed, as we see in Appendix~\ref{subsec:C2d}, this feat can be achieved. We find
\beq \D\mathfrak{a}\doteq\left(\frac{8\tilde g}{g}\right)\D a_{1g}+\D\tilde a_1,\eeq{beta8}
for some ambiguity $\tilde a_1$ spelled out in Eq.~\rf{2deramb}. Then one can choose
\beq \gt=\tf{1}{8}g,\eeq{choosegt} in order to fulfill the cocycle condition~\rf{cocycle3}. That is, Eq.~\rf{beta7}
takes the form:
\beq \D a_{1g}+\D a_{1\gt}+\D\tilde a_1\doteq-\G a_0,\eeq{2dervertex0}
where the vertex $a_0$ is given by
\beq a_0=ig\left(\bar\ps_{\m\a}R^{+\m\n\a\b}\ps_{\n\b}+\tf{1}{2}\bar{\pss}_\m\Rs^{\m\n}\pss_\n+\tf{1}{4}h_{\m\n}
\bar\uppsi_{\r\s\Vert\l}\,\g^{\m\r\s\a\b,\,\n\l\g}\,\uppsi_{\a\b\Vert\g}\right).\eeq{2dervertex1}

We emphasize that it is a unique linear combination in Eq.~\rf{1dera2}, with $\gt=\tf{1}{8}g$ being real valued, for which
the $a_2$ gets lifted to a vertex $a_0$ through the consistency cascade. The 2-derivative vertex is therefore unique. While
it simplifies in 4D as the last term in Eq.~\rf{2dervertex1} vanishes, the vertex is non-zero in any $D\geq4$.

%%%%%%%%%%%%%%%%%%%%%%%%%%%%%%%%%%%%%%%%%%%%%%%%%%%%%%%%%%%%%
%%%%%%%%%%%%%%%%%%%%%%%%%%%%%%%%%%%%%%%%%%%%%%%%%%%%%%%%%%%%%
\subsubsection{The 3-Derivative Vertex}\label{subsubsec:3der}
%%%%%%%%%%%%%%%%%%%%%%%%%%%%%%%%%%%%%%%%%%%%%%%%%%%%%%%%%%%%%
%%%%%%%%%%%%%%%%%%%%%%%%%%%%%%%%%%%%%%%%%%%%%%%%%%%%%%%%%%%%%

In this case, as we see from Eqs.~\rf{5halfa2-C} and~\rf{5halfa2}, there is just one candidate for $a_2$, namely
\beq a_2=-ig\,C^*_\l\,\bar{\upxi}_{\m\n}\g^{\l\m\n\a\b}\upxi_{\a\b}.\eeq{3der1}
Again, for future convenience, we wrote it with five $\g$-matrices, instead of just one as it appears in Eq.~\rf{5halfa2-C}.
The equivalence of the two forms is made explicit in Appendix~\ref{subsec:C3d}. Acting with $\D$ on Eq.~\rf{3der1} and
integrating by parts one evidently produces only $\G$-exact terms, thanks to the relations~\rf{gxicurl}. The corresponding
$a_1$ is thus easily seen to be
\beq a_1=-2igh^{*\s}_{\l}\left(\bar{\upxi}_{\m\n} \g^{\l\m\n\a\b}\uppsi_{\a\b\Vert\s}-\text{h.c.}\right)+\tilde{a}_1,\eeq{3der2}
for some ambiguity $\tilde a_1$ such that $\G\tilde a_1=0$. Now we address the problem of finding the lift to $a_0$. Acting on the above
expression with $\D$ again, one obtains the Einstein tensor, which can be written as
$G_{\l}^\s=2\de_{[\r}\mathfrak{h}^{\r\s\Vert}{}_{\l]}-\tf{1}{2}\d^\s_{\l}R$. Thus one ends up having
\beq \D a_1=-2ig\left(\de_{\r}\mathfrak{h}^{\r\s\Vert}{}_{\l}-\de_{\l}\mathfrak{h}^{\r\s\Vert}{}_{\r}-\tf{1}{2}\d^\s_{\l}R\right)
\left(\bar{\upxi}_{\m\n} \g^{\l\m\n\a\b}\uppsi_{\a\b\Vert\s}-\text{h.c.}\right)+\D\tilde{a}_1.\eeq{3der3}
The term proportional to the Ricci scalar is simply zero because of the Bianchi identity $\uppsi_{[\a\b\Vert\s]}=0$, while the
term containing $\de_\l$ is a total derivative, thanks again to the Bianchi identities $\de_{[\l}\bar\upxi_{\m\n]}=0$ and
$\de_{[\l}\bar\uppsi_{\a\b]\Vert\s}=0$, enforced by the presence of the antisymmetric 5-$\g$. Finally, the term containing
$\de_\r$ can be integrated by parts to give
\beq \D a_1\doteq 2ig\,\mathfrak{h}^{\r\s\Vert}{}_{\l}\,\left(\de_{\r}\bar{\upxi}_{\m\n}\,\g^{\l\m\n\a\b}\,\uppsi_{\a\b\Vert\s}
+\tf{1}{2}\bar{\upxi}_{\m\n}\,\g^{\l\m\n\a\b}\,\Ps_{\a\b|\r\s}-\text{h.c.}\right)+\D\tilde{a}_1.\eeq{3der4}
The first term in the parentheses and its hermitian conjugate combine into a $\G$-exact term modulo $d$, since the $\G$
variation of the graviton curl is zero up to a total derivative, again by the Bianchi identities $\de_{[\l}\bar\upxi_{\m\n]}=0$
and $\de_{[\l}\uppsi_{\a\b]\Vert\s}=0$. In the second term, on the other hand, one can pull a derivative out of the ghost-curl
and integrate by parts to obtain
\beq \D a_1+2ig\,\G\left(\mathfrak{h}^{\r\s\Vert}{}_{\l}\bar{\uppsi}_{\m\n\Vert\r}\,\g^{\l\m\n\a\b}\,\uppsi_{\a\b\Vert\s}\right)
\doteq igR_{\m\n\r\s}\left(\bar{\xi}_{\l}\,\g^{\l\m\n\a\b}\,\Ps_{\a\b|}{}^{\r\s}-\text{h.c.}\right)+\D\tilde{a}_1.\eeq{3der5}
As shown in Appendix~\ref{subsec:C3d}, the right-hand side can be rendered precisely $\G$-exact modulo $d$, with a choice of
the ambiguity, given by Eq.~\rf{3der6}. Then Eq.~\rf{3der5} reduces to
\beq \D a_1+2ig\,\G\left(\mathfrak{h}^{\r\s\Vert}{}_{\l}\bar{\uppsi}_{\m\n\Vert\r}\,\g^{\l\m\n\a\b}\,\uppsi_{\a\b\Vert\s}\right)
\doteq-ig\,\G\left(\mathfrak{h}^{\r\s\Vert\l}\bar{\uppsi}_{\m\n\Vert\g}\,\g^{\l\m\n\a\b,\,\r\s\g\d}\,\uppsi_{\a\b\Vert\d}
\right).\eeq{3der7}
The two $\G$-exact pieces then combine to have fulfilled the condition $\D a_1+\G a_0\doteq 0$, with
\beq a_0=ig\,\bar{\uppsi}_{\m\n\Vert}{}^\r\left(\mathfrak{h}^+_{\r\s\Vert\l}\,\g^{\l\m\n\a\b}+\g^{\l\m\n\a\b}\,
\mathfrak{h}^+_{\r\s\Vert\l}\right)\uppsi_{\a\b\Vert}{}^\s,\eeq{3der8}
where $\mathfrak{h}^{+\r\s\Vert\l}\equiv \mathfrak{h}^{\r\s\Vert\l}+\tf{1}{2}\g^{\r\s\a\b}\mathfrak{h}_{\a\b\Vert}{}^{\l}$.
The above 3-derivative vertex vanishes in $D=4$, and this fact is manifest from the presence of the antisymmetrized product
of five $\g$-matrices.

%%%%%%%%%%%%%%%%%%%%%%%%%%%%%%%%%%%%%%%%%%%%%%%%%%%
%%%%%%%%%%%%%%%%%%%%%%%%%%%%%%%%%%%%%%%%%%%%%%%%%%%
\subsection{Abelian Vertices}\label{subsec:Abelian}
%%%%%%%%%%%%%%%%%%%%%%%%%%%%%%%%%%%%%%%%%%%%%%%%%%%
%%%%%%%%%%%%%%%%%%%%%%%%%%%%%%%%%%%%%%%%%%%%%%%%%%%

Having exhausted all the nontrivial $a_2$'s, we are only left to consider vertices with trivial $a_2$. In this case, as we
show in Subsection~\ref{sec:2theorems} for generic spin, one can always choose to write a vertex as the graviton field
$h_{\m\n}$ contracted with a gauge-invariant\footnote{Gauge invariance of $T^{\m\n}$ is the whole point here; one can
always write a vertex as $a_0\approx T^{\mu\nu}h_{\mu\nu}$, but in general, e.g., for non-Abelian vertices, $T^{\mu\nu}$
will not be strictly gauge invariant.} current $T^{\m\n}$, \beq a_0=T^{\m\n} h_{\m\n},\qquad \G T^{\m\n}=0,\eeq{rs17}
where the divergence of the current is the $\D$ variation of a $\G$-closed object:
\beq \de_\n T^{\m\n}=\D M^\m ,\qquad \G M^\m=0.\eeq{rs18}

The gauge-invariant current $T^{\m\n}$ is a bilinear in the fermion fields, which cannot be $\D$-exact since otherwise the
vertex~\rf{rs17} would be trivial. This leaves us with considering only bilinears of the curvature $\Ps_{\m\n|\r\s}$.
Schematically, the current is of the form \beq T^{\m\n}=\bar\Ps^M\hat{\mathcal{O}}^{\m\n}{}_{MN}\Ps^N,\eeq{rs19} where $M,N$
are compound indices, and $\hat{\mathcal{O}}$ is an operator built from derivatives, $\g$-matrices and the metric tensor.
This immediately implies that an Abelian vertex will contain at least four derivatives$-$two from both curvatures with
$\hat{\mathcal{O}}$ containing no derivative.

To find the possible tensor structure of $\hat{\mathcal{O}}$, let us first note that we can forego contractions of any pair
of indices in the same curvature tensor since the result is always $\D$-exact, if not zero. It is sufficient to consider in
$\hat{\mathcal{O}}$ no more than one $\g$-matrix, which must carry either the $\m$-index or $\n$. To see this, notice that
if a $\g$-matrix carries one of the indices of the curvatures$-$any from the sets $M$ and $N-$one can use the Clifford
algebra to anticommute it past other possible $\g$-matrices to end up producing a $\g$-trace of the curvature, which is
$\D$-exact. This leaves us only with $\g^\m$ and $\g^\n$, which, however cannot appear simultaneously because their
symmetrization would eliminate them both. Similar reasonings rule out the appearance of the operator $\ds$, and therefore
of $\Box$, in $\hat{\mathcal{O}}$.

How many derivatives may $\hat{\mathcal{O}}$ contain? If it contains one derivative, there will be one $\g$-matrix carrying
either the index $\m$ or $\n$, say $\g^\m$. One can always choose the other index $\n$ to appear on the derivative under
consideration. In the only other nontrivial possibility, the latter index is contracted with, and therefore appears on, a
curvature on which the derivative must act. Then one can pull out the derivative $\de^\n$ by using the second Bianchi
identity and symmetry properties of the curvatures. Similarly, when $\hat{\mathcal{O}}$ contains more derivatives, one can
forego the appearance of the indices $\m$ and $\n$ on the curvatures. However, the number of derivatives cannot exceed two.
To see this, let us consider the possibility of having three derivatives or more:
\beq T^{\m\n}=\bar\Ps^M\overset\leftarrow{\de}_\r\hat{\mathcal{P}}^{\m\n}{}_{MN}\overset\rightarrow\de{}^\r\Ps^N,\nonumber
\eeq{qqq0}
where $\hat{\mathcal{P}}$ is a 1- or higher-derivative operator.
Then one can use the so-called 3-box rule: $2\de_\r X\de^\r Y=\Box(XY)-X\Box Y-Y\Box X$, integrate by parts, and drop $\D$-exact terms
to write
\beq a_0\approx\Box h_{\m\n}\left(\tf{1}{2}\bar\Ps^M\hat{\mathcal{P}}^{\m\n}{}_{MN}\Ps^N\right)\approx\left(\tf{1}{2}\de_\m
h'-\de\cdot h_\m\right)\de_\n\left(\bar\Ps^M\hat{\mathcal{P}}^{\m\n}{}_{MN}\Ps^N\right),\nonumber\eeq{qqq1}
where the last equivalence is due to the fact that $R_{\m\n}\equiv\Box h_{\m\n}-2\de_{(\m}\de\cdot h_{\n)}+\de_\m\de_\n h'$
is a $\D\text{-exact}$ quantity. Therefore, the vertex is trivial since the divergence of the fermion bilinear is $\D$-exact.
The latter fact originates from $\de_\n T^{\m\n}=\D M^\m$, and that the divergence is blind to the presence of the extra
derivatives $\overset\leftarrow{\de}_\r\overset\rightarrow\de{}^\r$ in $T^{\m\n}$. On the other hand, if the extra derivatives
carry any indices belonging to the sets $M$ and $N$, one must keep in mind that a divergence of the curvature is $\D$-exact.
Given the hermiticity of $T^{\m\n}$, the commutativity of covariant derivatives, the antisymmetry of paired indices in and the
Bianchi identities obeyed by the curvature, it is easy to convince ourselves that this vertex is always equivalent to the
previous one, which we already ruled out. This proves our claim that $\hat{\mathcal{O}}$ may contain at most two derivatives.
This sets an upper bound of six on the number of derivatives in $T^{\m\n}$, and therefore also in the vertex $a_0$.

%%%%%%%%%%%%%%%%%%%%%%%%%%%%%%%%%%%%%%%%%%%%%%%%%%%%%%%%%%%%%
\subsubsection{The 4-Derivative Vertex}\label{subsubsec:4der}
%%%%%%%%%%%%%%%%%%%%%%%%%%%%%%%%%%%%%%%%%%%%%%%%%%%%%%%%%%%%%

When the operator $\hat{\mathcal{O}}$ in Eq.~\rf{rs19} does not contain any derivative, the corresponding vertex~\rf{rs17}
is a 4-derivative one. The generic form of the current is
\beq T^{\m\n}=ig\left(\bar\Ps^{(\m}{}_{\l|\a\b}\Ps^{\n)\l|\a\b}+\a\h^{\m\n}\bar\Ps_{\r\s|\a\b}\Ps^{\r\s|\a\b}\right),\eeq{4der0}
where the parameter $\a$ is to be fixed by requiring that $\de_\n T^{\m\n}$ be $\D$-exact. Now the divergence of Eq.~\rf{4der0}
contains some nontrivial pieces in $H(\D)$, given by
\beq \de_\n T^{\m\n}=\D M^\m+ig\left(\tf{1}{2}\bar\Ps^{\n\l|\a\b}\de_\n\Ps^\m{}_{\l|\a\b}+\a\bar\Ps^{\r\s|\a\b}
\de^\m\Ps_{\r\s|\a\b}-\hc\right).\eeq{4der1}
By using the Bianchi identity $\de_{[\n}\Ps^\m{}_{\l]|\a\b}=\tf{1}{2}\de^\m\Ps_{\n\l|\a\b}$, the first term in the parentheses
is rendered the same as the second one. These terms cancel each other if we set $\a=-\tf{1}{4}$. Thus, there is just one
4-derivative vertex, given by
\beq a_0~=~ig\left(h_{\m\n}-\tf{1}{4}\h_{\m\n}h'\right)\bar\Ps^\m{}_{\l|\a\b}\Ps^{\n\l|\a\b}~\approx~-\tf{i}{2}g\left(h_{\m\n}
-\tf{1}{4}\h_{\m\n}h'\right)\bar\Ps^\m{}_{\l|\r\s}\g^{\r\s\a\b}\Ps^{\n\l|}{}_{\a\b},\eeq{4der2}
where the last equivalent form owes its existence to the identity~\rf{C.3}, which can be rewritten as
$\h^{\r\s|\a\b}=-\tf{1}{2}\g^{\r\s\a\b}+\tf{1}{2}\g^{\r\s}\g^{\a\b}-2\g^{[\r}\h^{\s][\a}\g^{\b]}$, and to the
EoMs~\rf{Fronsdal1curl} and~\rf{eom5/24}.

Now let us compute the quantity $\D M^\m=\de_\n T^{\m\n}$ from Eq.~\rf{4der2}. One gets
\beq \D M^\m=-\tf{i}{4}g\,\bar\Ps^\m{}_{\l|\r\s}\g^{\r\s\a\b}\de_\n\Ps^{\n\l|}{}_{\a\b}+\hc\,.\eeq{4der4}
Now one can use the identity~\rf{eom5/26}, for the divergence of the curvature, to obtain
\beq \D M^\m=-\tf{1}{2}g\,\bar\Ps^{\m\l|}{}_{\r\s}\g^{\r\s\a\b}\ds\,\de_{[\a}\mathcal{S}_{\b]\l}+\tf{1}{4}g\,
\bar\Ps^{\m\l|}{}_{\r\s}\g^{\r\s\a\b}\de_\l\de_{[\a}\displaystyle{\not{\!\mathcal{S}}}_{\b]}+\hc\,.\eeq{4der5}
In the first term on the right-hand side, one can use $\g^{\r\s\a\b}\ds=\left(2\g^{\r\s\a\b\t}-\g^\t\g^{\r\s\a\b}\right)\de_\t$,
and then integrate by parts w.r.t. $\de_\t$ noticing that the 5-$\g$ piece is killed by Bianchi identity. In the second
term, on the other hand, one can integrate by parts w.r.t. $\de_\l$. The result is
\bea \D M^\m&=&\tf{1}{2}g\,\de_\t\left(\bar\Ps^{\m\l|}{}_{\r\s}\g^\t\g^{\r\s\a\b}\de_{[\a}\mathcal{S}_{\b]\l}\right)+\tf{1}{4}
g\,\de_\l\left(\bar\Ps^{\m\l|}{}_{\r\s}\g^{\r\s\a\b}\de_{[\a}\displaystyle{\not{\!\mathcal{S}}}_{\b]}\right)\nonumber\\
&&-\tf{1}{2}g\,\bar\Ps^{\m\l|}{}_{\r\s}\overset\leftarrow{\ds}\,\g^{\r\s\a\b}\de_{[\a}\mathcal{S}_{\b]\l}-\tf{1}{4}g\,
\bar\Ps^{\m\l|}{}_{\r\s}\overset\leftarrow{\de}_\l\,\g^{\r\s\a\b}\de_{[\a}\displaystyle{\not{\!\mathcal{S}}}_{\b]}+\hc\,.
\eea{4der6}
The first line on the right-hand side is a double divergence, because one can pull out the $\de_\a$ from the Fronsdal tensor,
and make it a total derivative by using the Bianchi identities. That is, the first line plus its hermitian conjugate reduces
to the form $\de_\a\de_\t\mathcal Y_1^{\m\a\t}$, where
\beq \mathcal Y_1^{\m\a\t}=\tf{1}{2}g\left(\bar\Ps^{\m\l|}{}_{\r\s}\g^\t\g^{\r\s\a\b}\mathcal{S}_{\b\l}+\tf{1}{2}
\bar\Ps^{\m\t|}{}_{\r\s}\g^{\r\s\a\b}\displaystyle{\not{\!\mathcal{S}}}_\b+\hc\right),\eeq{4der7}
which is both $\G$-closed and $\D$-exact. On the other hand, the second line of Eq.~\rf{4der6} contains bilinears in the
Fronsdal tensor by virtue of the EoMs~\rf{eom5/25} and~\rf{eom5/26}. The first piece contains the double curl,
$\de^{[\m}\de_{[\r}\bar{\mathcal S}_{\s]}{}^{\l]}$, while the second one includes
$\de_{[\r}\bar{\mathcal S}_{\s]}{}^\m\overset\leftarrow\ds$. In the former of these, one pulls out $\de_\m$ to
integrate by parts, while in the latter one uses $\overset\leftarrow\ds\,\g^{\r\s\a\b}=\overset\leftarrow\de_\t\left(
2\g^{\r\s\a\b\t}-\g^{\r\s\a\b}\g^\t\right)$ and then integrate by parts w.r.t. $\de_\t$. The last step produces
$\ds\,\de_{[\a}\displaystyle{\not{\!\mathcal{S}}}_{\b]}$, which then can be replaced, thanks to identity~\rf{x17},
by $2\de^\l\de_{[\a}\mathcal{S}_{\b]\l}$. The same step also gives a total derivative:
$\de_\t\left(\de_{[\r}\bar{\mathcal S}_{\s]}{}^\m\g^{\r\s\a\b}\g^\t\de_{[\a}\displaystyle{\not{\!\mathcal{S}}}_{\b]}\right)$,
which can be turned into a double divergence by pulling out $\de_\a$ and integrating by parts. When hermitian conjugates
are taken into account, the end result is that the second line of Eq.~\rf{4der6} reduces to the form
$\de_\n\mathcal X^{(\m\n)}+\de_\a\de_\t\mathcal Y_2^{\m\a\t}$, where $\mathcal X$ and $\mathcal Y_2$ are both
$\G$-closed and $\D$-exact:
\bea \mathcal Y_2^{\m\a\t}&=&-\tf{i}{2}g\left(\de_{[\r}\bar{\mathcal S}_{\s]}{}^\m\g^{\r\s\a\b}\g^\t\displaystyle
{\not{\!\mathcal{S}}}_\b-\hc\right),\label{4der8}\\ \mathcal X^{(\m\n)}&=&-\tf{i}{4}g\left(\de_{[\r}\bar{\mathcal S}_{\s]}
{}^\m\g^{\r\s\a\b}\de_{[\a}\mathcal {S}_{\b]}{}^\n+\de_{[\r}\bar{\mathcal S}_{\s]}{}^\n\g^{\r\s\a\b}\de_{[\a}\mathcal {S}_{\b]}{}^\m\right)\nonumber\\&&+\tf{i}{4}g\,\h^{\m\n}\left(\de_{[\r}\bar{\mathcal S}_{\s]}{}^\l\g^{\r\s\a\b}
\de_{[\a}\mathcal {S}_{\b]\l}+\tf{1}{4}\de_{[\r}\bar{\displaystyle{\not{\!\mathcal{S}}}}_{\s]}\g^{\r\s\a\b}
\de_{[\a}\displaystyle{\not{\!\mathcal{S}}}_{\b]}\right).\eea{4der9}
Thus, we have shown that $\D M^\m$ can be rewritten as
\beq\D M^\m=\de_\n\mathcal X^{(\m\n)}+\de_\a\de_\t\left(\mathcal Y_1^{\m\a\t}+\mathcal Y_2^{\m\a\t}\right).\eeq{4der10}
This, along with Eqs.~\rf{4der7}--\rf{4der9}, fulfills the sufficient condition~\rf{abel0} for the triviality of $a_1$.
That is, the vertex does not actually deform the gauge transformations: one can make it strictly gauge-invariant modulo $d$,
by adding $\D$-exact terms spelled out in Eq.~\rf{abel3}.

Although not manifest, this vertex actually vanishes in 4D. In fact, one can find the following form of the vertex:
\beq a_0\approx-\tf{i}{8}g\,h_{\m\n}\bar\Ps_{\r\s|\t\l}\,\g^{\m\r\s\a\b,\,\n\t\g}\,\Ps_{\a\b|\g}{}^\l,\eeq{4der11}
which makes the triviality in $D=4$ manifest. To see that this is indeed equivalent to the vertex~\rf{4der2}, let is use
the $\g$-matrix identity~\rf{C.18} in the vertex~\rf{4der11} to break it into terms containing only antisymmetric products
of six $\g$-matrices or two. The former kind of terms all vanish because of either the Bianchi identities or the symmetry
in the indices carried by the graviton. On the other hand, the terms containing two $\g$-matrices are actually equivalent
to terms containing none. This is because the symmetry in the graviton indices requires that at least one $\g$-matrix be
contracted with a spin-$\tf{5}{2}$ curvature; then the Clifford algebra gives a $\g$-trace of the curvature,
which is $\D$-exact. Thus we get
\beq a_0\approx-\tf{i}{8}g h_\m{}^\n\,\bar\Ps_{\r\s|\l}{}^\t\left[12\,\d^{\b\r\s}_{\n\t\g}\,\h^{\m\a}+24\,\d^{\m\r\b}_{\n\t\g}
\,\h^{\s\a}-12\,\d^{\s\a\b}_{\n\t\g}\,\h^{\m\r}\right]\Ps_{\a\b|}{}^{\l\g}.\nonumber\eeq{4der12}
Having gotten rid of $\g$-matrices, it is now straightforward to carry the computation. The number of possible terms are
greatly reduced by the symmetry properties of the associated fields and curvatures. One can also drop traces of the curvatures
since they are $\D$-exact. Thus, one ends up having the first form of the vertex presented in Eq.~\rf{4der2}.

%%%%%%%%%%%%%%%%%%%%%%%%%%%%%%%%%%%%%%%%%%%%%%%%%%%%%%%%%%%%%
\subsubsection{The 5-derivative vertex}\label{subsubsec:5der}
%%%%%%%%%%%%%%%%%%%%%%%%%%%%%%%%%%%%%%%%%%%%%%%%%%%%%%%%%%%%%

When the vertex contains five derivatives, the operator $\hat{\mathcal{O}}$ in Eq.~\rf{rs19} includes one.
As we discussed already, the form of $\hat{\mathcal O}$ is much restricted. Indeed, we have just one possibility:
\beq T^{\m\n}=ig\,\bar\Ps^{\r\s|\a\b}\g^{(\m}\overset\leftrightarrow\de{}^{\n)}\Ps_{\r\s|\a\b},\eeq{5der0}
where the operator $\overset\leftrightarrow\de_\m\equiv\overset\rightarrow\de_\m-\overset\leftarrow\de_\m$
plays a crucial role in eliminating from $\de_\n T^{\m\n}$ terms that are not $\D$-exact. The vertex is given,
by Eq.~\rf{rs17}, as
\beq a_0~=~ig h_{\m\n}\bar\Ps^{\r\s|\a\b}\g^{(\m}\overset\leftrightarrow\de{}^{\n)}\Ps_{\r\s|\a\b}~\approx~
-ig\,\mathfrak{h}_{\m\n\Vert\l}\bar\Ps^\m{}_{\t|\r\s}\,\g^\l\,\Ps^{\n\t|\r\s}.\eeq{5der1}
To see the equivalence of the second form, let us remove therein any derivative on the graviton field by
integrating it by parts. This gives a derivative of the spin-$\tf{5}{2}$ curvatures: the divergence is $\D$-exact,
while in the gradient one can use the second Bianchi identity and the symmetry properties of the curvatures to
pull out a derivative with an index of the graviton field. The equivalence of the vertices then follows immediately.

We can write the second equivalent form as~~$\tf{1}{2}\mathfrak{h}_{\m\n\Vert\l}\bar\Ps^{\m\t|}{}_{\r\s}\left(\h^{\r\s|\a\b}
\g^\l+\g^\l\h^{\r\s|\a\b}\right)\Ps^\n{}_{\t|\a\b}$. Then the identity
$\h^{\r\s|\a\b}=-\tf{1}{2}\g^{\r\s\a\b}+\tf{1}{2}\g^{\r\s}\g^{\a\b}-2\g^{[\r}\h^{\s][\a}\g^{\b]}$ helps us drop
some $\D$-exact pieces, thanks to Eqs.~\rf{Fronsdal1curl}--\rf{eom5/24}, to be left with
$\tf{1}{2}\left(\g^{\r\s\a\b}\g^\l+\g^\l\g^{\r\s\a\b}\right)=\g^{\l\r\s\a\b}$. Therefore, we have another
equivalent form of the vertex:
\beq a_0\approx\tf{i}{2}g\,\mathfrak{h}_{\m\n\Vert\l}\bar\Ps^{\m\t|}{}_{\r\s}\,\g^{\l\r\s\a\b}\,\Ps^\n{}_{\t|\a\b}.\eeq{5der2}
The virtue of this form is twofold. First, the presence of an antisymmetric product of five $\g$-matrices manifestly
renders this vertex trivial in $D=4$. Second, because of the Bianchi identities, the gauge variation of the vertex is
just a total derivative, which means that it does not deform the gauge transformations.

%%%%%%%%%%%%%%%%%%%%%%%%%%%%%%%%%%%%%%%%%%%%%%%%%%%%%%%%%%%%%
\subsubsection{The 6-Derivative Vertex}\label{subsubsec:6der}
%%%%%%%%%%%%%%%%%%%%%%%%%%%%%%%%%%%%%%%%%%%%%%%%%%%%%%%%%%%%%

There is a unique 6-derivative hermitian current whose divergence is $\D$-exact. It reads
\beq T^{\m\n}=ig\,\bar\Ps^{\r\s|\a\b}\left(\overset\rightarrow{\de}{}^\m\overset\rightarrow{\de}{}^\n+\overset\leftarrow{\de}{}^\m
\overset\leftarrow{\de}{}^\n-\h^{\m\n}\overset\leftarrow{\de}{}^\l\overset\rightarrow{\de}{}_\l\right)\Ps_{\r\s|\a\b}.\eeq{6der0}
While the vertex is simply given by $T^{\m\n}h_{\m\n}$, one can also cast it into a ``geometrical'' form that involves the product
of all three curvatures:
\beq a_0\approx ig R_{\m\n\r\s}\bar\Ps^{\r\s|\a\b} \Ps_{\a\b}{}^{\m\n}.\eeq{6der1}
This form is strictly gauge invariant, and the vertex exists in all $D\geq4$. To see the equivalence of the two forms of the
vertex, let us remove in the vertex~\rf{6der1} all the derivatives from the graviton field, by integrations by parts. Dropping
divergences of the spin-$\tf{5}{2}$ curvature, that are $\D$-exact, we arrive at
\beq a_0\approx 4ig h_{\m\n}\bar\Ps^{\m\a|\r\s}\overset\leftarrow{\de}{}^\b\overset\rightarrow{\de}_\a \Ps^\n{}_{\b|\r\s}
\approx 4ig h_{\m\n}\left(-\bar\Ps^{\a\b|\r\s}\overset\leftarrow{\de}{}^\m+\bar\Ps^{\m\b|\r\s}\overset\leftarrow{\de}{}^\a\right)
\overset\rightarrow{\de}_\a \Ps^\n{}_{\b|\r\s},\nonumber\eeq{6der2}
where the second equivalence results from the Bianchi identity. The first term in the parentheses imposes the Bianchi identity
$\de_{[\a}\Ps^\n{}_{\b]|\r\s}=\tf{1}{2}\de^\n\Ps_{\a\b|\r\s}$, whereas the second term enables us to use the 3-box rule:
$2\de^\a X\de_\a Y=\Box(XY)-X\Box Y-Y\Box X$, so that we can drop $\D$-exact terms, like $\Box\Ps_{\m\b|\r\s}$, and integrate
by parts to obtain
\beq a_0\approx -2ig h_{\m\n}\bar\Ps^{\a\b|\r\s}\overset\leftarrow{\de}{}^\m\overset\rightarrow{\de}{}^\n \Ps^{\a\b|\r\s}
+2ig\,\Box h_{\m\n}\bar\Ps^{\m\b|\r\s}\Ps^\n{}_{\b|\r\s}.\nonumber\eeq{6der3}
Now, let us replace $\Box h_{\m\n}$ by $2\de_{(\m}\de\cdot h_{\n)}-\de_\m\de_\n h'$, since their difference is
$R_{\m\n}=\D\text{-exact}$. In the resulting expression, let us remove all the derivatives from the graviton field to get
\beq a_0\approx-2ig h_{\m\n}\bar\Ps^{\a\b|\r\s}\overset\leftarrow{\de}{}^\m\overset\rightarrow{\de}{}^\n \Ps^{\a\b|\r\s}
+2ig\left(h_{\m\n}-\tf{1}{2}\h_{\m\n} h'\right)\de^\n\left(\bar\Ps^{\l\b|\r\s}\overset\rightarrow{\de}{}_\l\Ps^\m{}_{\b|\r\s}
-\hc\right).\nonumber\eeq{6der4}
In the second term, we can again use the Bianchi identity $\de_{[\l}\Ps^\m{}_{\b]|\r\s}=\tf{1}{2}\de^\m\Ps_{\l\b|\r\s}$
to find that some of the resulting pieces cancel the first term. The remaining pieces add to
the form $T^{\m\n}h_{\m\n}$, with $T^{\m\n}$ given precisely by Eq.~\rf{6der0}. So the vertices are equivalent.

%%%%%%%%%%%%%%%%%%%%%%%%%%%%%%%%%%%%%%%%%%%%%%%%%%%%%%%%%%%%%%%%%%%%%%%%%%%%%%%%%%%%%%%%%%%%%%%%%%%%%%%%%%%%%%
%%%%%%%%%%%%%%%%%%%%%%%%%%%%%%%%%%%%%%%%%%%%%%%%%%%%%%%%%%%%%%%%%%%%%%%%%%%%%%%%%%%%%%%%%%%%%%%%%%%%%%%%%%%%%%
\section{Arbitrary Spin: $\textbf{\textit s}\,\bf{=}\,\textbf{\textit n}+\bf{\tf{1}{2}}$}\label{sec:arbitrary}
%%%%%%%%%%%%%%%%%%%%%%%%%%%%%%%%%%%%%%%%%%%%%%%%%%%%%%%%%%%%%%%%%%%%%%%%%%%%%%%%%%%%%%%%%%%%%%%%%%%%%%%%%%%%%%
%%%%%%%%%%%%%%%%%%%%%%%%%%%%%%%%%%%%%%%%%%%%%%%%%%%%%%%%%%%%%%%%%%%%%%%%%%%%%%%%%%%%%%%%%%%%%%%%%%%%%%%%%%%%%%

The sets of fields and antifields for the arbitrary-spin case are given by
\beq \Phi^A=\{h_{\m\n}, C_\m, \ps_{\m_1...\m_n}, \xi_{\m_1...\m_{n-1}}\},\qquad \Phi^*_{A}=\{h^{*\m\n}, C^{*\m},
\bar{\ps}^{*\m_1...\m_n}, \bar{\xi}^{*\m_1...\m_{n-1}}\}.\eeq{arb1}
For $n>2$, there is a triple $\g$-trace constraint on the field and antifield, i.e.,
\beq \pss'_{\m_1...\m_{n-3}}=0,\qquad \bar\pss^{*\prime}_{\m_1...\m_{n-3}}=0,\eeq{arb2}
The rank-($n-1$) fermionic ghost and its antighost are $\g$-traceless as usual:
\beq \displaystyle{\not{\!\xi}}_{\m_1...\m_{n-2}}=0,\qquad \displaystyle{\not{\!\bar{\xi}}}^{\,*}_{\m_1...\m_{n-2}}=0.\eeq{arb3}

The spin-$s$ Lagrangian EoMs are given by the rank-$n$ tensor-spinor $\mathcal R_{\m_1...\m_n}$, which is an arbitrary-spin
generalization of~\rf{EoM5/2}, and is related to the Fronsdal tensor as follows:
\beq \mathcal R_{\m_1...\m_n}=\mathcal S_{\m_1...\m_n}-\tf{1}{2}n\,\g_{(\m_1}\displaystyle\not{\!\mathcal S}_{\m_2...\m_n)}
-\tf{1}{4}n(n-1)\,\h_{(\m_1\m_2}\mathcal S^\prime_{\m_3...\m_n)}.\eeq{arb4}

While an account of the cohomology of $\G$ is given in Appendix~\ref{sec:cohomology}, below we spell out some important properties
of the various fields and antifields.
\begin{table}[ht]
\caption{Properties of the Various Fields \& Antifields ($n\;\text{arbitrary}$)}
\vspace{6pt}
\centering
\begin{tabular}{c c c c c c c}
\hline\hline
$Z$ &$\G(Z)$~~~&~~~$\D(Z)$~~~&$pgh(Z)$ &$agh(Z)$ &$gh(Z)$ &$\epsilon(Z)$\\ [0.5ex]
\hline
$h_{\m\n}$ & $2\de_{(\m} C_{\n)}$ & 0 & 0 & 0 & 0 & 0\\
$C$ & 0 & 0 & 1 & 0 & 1 & 1\\
$h^{*\m\n}$ & 0 & $G^{\m\n}$ & 0 & 1 & $-1$ & 1\\
$C^{*\m}$ & 0 & $-2\de_\n h^{*\m\n}$ & 0 & 2 & $-2$ & 0\\ \hline
$\ps_{\m_1...\m_n}$ & $n\de_{(\m_1}\xi_{\m_2...\m_n)}$ & 0 & 0 & 0 & 0 & 1\\
$\xi_{\m_1...\m_{n-1}}$ & 0 & 0 & 1 & 0 & 1 & 0\\
$\bar{\ps}^{*\m_1...\m_n}$ & 0 & $\bar{\mathcal R}^{\m_1...\m_n}$ & 0 & 1 & $-1$ & 0\\
$\bar{\xi}^{*\m_1...\m_{n-1}}$ & 0 & $2\de_{\m_n}\bar{\chi}^{*\m_1...\m_n}$ & 0 & 2 & $-2$ & 1\\
\hline\hline
\end{tabular}
\end{table}
\vspace{6pt}

Note that the antifield $\bar{\chi}^{*\m_1...\m_n}$ is given by Eqs.~\rf{Norther-n2}--\rf{Noether-n3}. With the above table,
it is easy to construct the BRST-closed free master action for the arbitrary-spin case:
\bea S_0&=&\int d^Dx\left[G^{\m\n}h_{\m\n}+\tf{1}{2}\left(\bar{\mathcal R}^{\m_1...\m_n}\ps_{\m_1...\m_n}-\bar{\ps}_{\m_1...\m_n}
\mathcal{R}^{\m_1...\m_n}\right)\right]\nonumber\\&&+\int d^Dx\left[-2h^{*\m\n}\de_\m C_\n+\tf{n}{2}(\bar{\ps}^{*\m_1...\m_n}
\de_{\m_1}\xi_{\m_2...\m_n}-\de_{\m_1}\bar{\xi}_{\m_2...\m_n}\ps^{*\m_1...\m_n})\right].\eea{FMA-n}

Now we are ready to construct the $2-s-s$ cubic vertices. Having worked out the spin $\tf{5}{2}$ case as a prototypical example,
our job has become easy, since many of the statements made for spin $\tf{5}{2}$ go verbatim for arbitrary spin.

%%%%%%%%%%%%%%%%%%%%%%%%%%%%%%%%%%%%%%%%%%%%%%%%%%%%%
%%%%%%%%%%%%%%%%%%%%%%%%%%%%%%%%%%%%%%%%%%%%%%%%%%%%%
\subsection{Non-Abelian Vertices}\label{sec:1theorem}
%%%%%%%%%%%%%%%%%%%%%%%%%%%%%%%%%%%%%%%%%%%%%%%%%%%%%
%%%%%%%%%%%%%%%%%%%%%%%%%%%%%%%%%%%%%%%%%%%%%%%%%%%%%

Let us recall that any $a_2$ consists of two ghost fields and a single antighost, and that the latter can be chosen
to be undifferentiated without loss of generality. As explained in Appendix~\ref{sec:cohomology}, a single
derivative acting on the ghost $C_\m$ can be realized as a 1-curl $\mathfrak C_{\m\n}$ modulo irrelevant
$\G$-exact terms, while two or more derivatives are never nontrivial. For the fermionic ghost $\xi_{\m_1...\m_{n-1}}$,
on the other hand, one can choose any $m$-curl: $\upxi^{(m)}_{\m_1\n_1|...|\m_m\n_\m\Vert\m_{m+1}...\m_{n-1}}$ with
$m=1,2,...,n-1$, and more than $n-1$ derivatives give $\G$-exact terms. Clearly, a nontrivial $a_2$ cannot contain
more than $2n-2$ derivatives. This sets an upper bound of $2n-1$ on the number of derivatives in a non-Abelian vertex
given the actions of $\G$ and $\D$ on various (anti)fields and the consistency cascade~\rf{cocycle1}--\rf{cocycle3}.

Again, all nontrivial $a_2$'s fall into two subsets: Subset-1 contains the bosonic antighost $C^{*\m}$, and subset-2 the
fermionic one $\xi^{*\m_1...\m_{n-1}}$. Subset-1 has the form: $a_2=C^{*\m} X_\m$,
where $X_\m$ is some bilinear in the fermionic ghost-curls. Then we have: $\D a_2\doteq 2h^{*\m\n}\de_{(\m}X_{\n)}$,
which must be $\G$-exact modulo $d$ if the cocycle condition~\rf{cocycle2} is to be satisfied. Because $\G$ does not
act on the antifields, a functional derivative w.r.t. $h^{*\m\n}$ gives
\beq \de_{(\m}X_{\n)}=\G\text{-exact}.\eeq{xarb2}
Now, the symmetrized derivative of $X_\m$ can be schematically written as
\bea \de X&\sim&\de\left[\bar\upxi^{(m_1)}\upxi^{(m_2)}\pm\bar\upxi^{(m_2)}\upxi^{(m_1)}\right]\nonumber\\
&\sim&\G\text{-exact}+\bar\upxi^{(m_1+1)}\upxi^{(m_2)}+\bar\upxi^{(m_1)}\upxi^{(m_2+1)}\pm\bar\upxi^{(m_2+1)}
\upxi^{(m_1)}\pm\bar\upxi^{(m_2)}\upxi^{(m_1+1)}.\eea{xarb3}
When $m_1$ and $m_2$ are equal, we have the plus sign for a nonzero $X$, and nontrivial elements of $H(\G)$ are absent only
when $m_1=m_2=n-1$. When they are unequal, let us take $m_1>m_2$, and then $\de X$ is $\G$-exact with the minus sign if
$m_1=m_2+1=n-1$. The only $a_2$'s that pass the condition~\rf{cocycle2} thus contain $2n-3$ and $2n-2$ derivatives. More explicitly,
\beq a_2=\begin{cases}p=2n-2:\quad ig\,C^{*\m}\left[\bar\upxi^{(n-1)}_{\m\,\cdots}\upxi^{(n-2)\,\cdots}
-\bar\upxi^{(n-2)\,\cdots}\upxi^{(n-1)}_{\m\,\cdots}\right]\\
p=2n-1:\quad ig\,C^{*\m}\bar\upxi^{(n-1)\,\cdots}\g_\m\upxi^{(n-1)}_{\,\cdots},\end{cases}\eeq{arba2-C}
where the ellipses mean contracted indices. This is very similar to the spin-$\tf{5}{2}$ case.

Subset-2, on the other hand, has the (undifferentiated) fermionic antighost. In this case, the $a_2$'s have the form:
$a_2=\bar\xi^{*\m_1...\m_{n-1}}Y_{\m_1...\m_{n-1}}+\hc$\,. Symmetry is imposed in the indices of $Y$, which comprises both of
the ghosts and curls thereof. Following the same logic as presented for spin $\tf{5}{2}$, it is clear that $a_2$ can
contain at most two derivatives: one in $\mathfrak C_{\m\n}$ and the other in the 1-curl
$\upxi^{(1)}_{\m_1\n_1\Vert\n_2...\n_{n-1}}$, as higher-curls of the latter are incompatible with the symmetry of the
indices. At this point the possibilities (all to be ruled out) are:
\beq a_2=\begin{cases}p=0:\quad g\,\bar\xi^{*\m\,\cdots}\g^\a\xi_{\m\,\cdots} C_\a+\hc\\
p=1:\quad g\,\bar\xi^{*\m\,\cdots}\left(\xi^\n{}_{\cdots}\mathfrak C_{\m\n}+\a_1\upxi^{(1)}_{\m\n\Vert\cdots}C^\n
+\a_2\g^{\a\b}\xi_{\m\,\cdots}\mathfrak C_{\a\b}\right)+\hc\\
p=2:\quad g\,\bar\xi^{*\m\,\cdots}\g^\a\upxi^\b{}_{\m\Vert\cdots}\mathfrak{C}_{\a\b}+\hc\,.\end{cases}\eeq{arba2xi}
However, one can derive quite similarly a counterpart of condition~\rf{xarb2}, namely
\beq \de_{(\m_1}Y_{\m_2...\m_n)}=\G\text{-exact}.\eeq{xarb4}
When $n>2$, it is impossible for any element in the list~\rf{arba2xi} to fulfill this condition because $\de\,Y$ will always
contain nontrivial elements of $H(\G)$. This rules out all of them.

%%%%%%%%%%%%%%%%%%%%%%%%%%%%%%%%%%%%%%%%%%%%%%%%%%%%%%%%%%%%%%%%%%%%%%%%%%%%%%%%%%%%%
\subsubsection*{The $\textbf{(2\textit{n}-2)}$-Derivative Vertex}\label{sec:2nminus2}
%%%%%%%%%%%%%%%%%%%%%%%%%%%%%%%%%%%%%%%%%%%%%%%%%%%%%%%%%%%%%%%%%%%%%%%%%%%%%%%%%%%%%

In this case, one can go along the same line as the 2-derivative spin-$\tf{5}{2}$ vertex. To make the steps go verbatim
we add a trivial term to the first element of~\rf{arba2-C}, and write
\beq a_2=ig\,C^{*\m}\left[\bar\upxi^{(n-1)}_{\m\,\cdots}\upxi^{(n-2)\,\cdots}-\bar\upxi^{(n-2)\,\cdots}
\upxi^{(n-1)}_{\m\,\cdots}\right]+\tf{1}{8}g\,\mathfrak{C}_{\m\n}\left[\bar\upxi^{*(n-2)}_{\cdots\Vert\r}\g^{\m\n\r\a\b}
\upxi^{(n-1)\cdots\Vert}{}_{\a\b}-\hc\right],\eeq{xarb5}
which looks quite similar to the spin-$\tf{5}{2}$ counterpart~\rf{1dera2}, given the relation~\rf{choosegt}. To obtain
the vertex, one can simply redo the steps of Subsection~\ref{subsubsec:2der}. One finds,
\bea a_0&=&ig\left[\,\bar\uppsi^{(n-2)}_{\cdots\Vert\,\m\a}R^{+\m\n\a\b}\uppsi^{(n-2)\cdots\Vert}{}_{\n\b}+\tf{1}{2}
\bar{\displaystyle{\not{\!\uppsi}}}^{(n-2)}_{\cdots\Vert\,\m}\Rs^{\m\n}\displaystyle{\not{\!\uppsi}}^{(n-2)\cdots\Vert}
{}_\n\,\right]\nonumber\\&&~~~~~~~~~~~~~~~~~~~~~~~~+\tf{i}{4}g\,h_{\m\n}\,\bar\uppsi^{(n-1)}_{\cdots\r\s\Vert\,\l}\,
\g^{\m\r\s\a\b,\,\n\l\g}\,\uppsi^{(n-1)\cdots}{}_{\a\b\Vert\,\g}\,,\eea{xarb6}
as the desired non-Abelian $2-s-s$ vertex containing $2n-2$ derivatives. Again, let us notice the striking similarity
with its spin-$\tf{5}{2}$ counterpart~\rf{2dervertex1}.

%%%%%%%%%%%%%%%%%%%%%%%%%%%%%%%%%%%%%%%%%%%%%%%%%%%%%%%%%%%%%%%%%%%%%%%%%%%%%%%%%%%%%
\subsubsection*{The $\textbf{(2\textit{n}-1)}$-Derivative Vertex}\label{sec:2nminus1}
%%%%%%%%%%%%%%%%%%%%%%%%%%%%%%%%%%%%%%%%%%%%%%%%%%%%%%%%%%%%%%%%%%%%%%%%%%%%%%%%%%%%%

Here one starts with the second element of~\rf{arba2-C} as the starting point. We use five $\g$-matrices, instead of one,
to have a direct generalization of Eq.~\rf{3der1}:
\beq a_2=-ig\,C^*_\l\bar\upxi^{(n-1)}_{\,\cdots|\,\m\n}\,\g^{\l\m\n\a\b}\,\upxi^{(n-1)\cdots|}{}_{\,\a\b}.\eeq{xarb7}
One can proceed in the same way as in Subsection~\ref{subsubsec:3der} to find:

\beq a_0=ig\,\bar{\uppsi}^{(n-1)}_{\,\cdots\,\m\n\,\Vert}{}^{\,\r}\left(\mathfrak{h}^+_{\r\s\Vert\l}\,\g^{\l\m\n\a\b}+\g^{\l\m\n\a\b}\,
\mathfrak{h}^+_{\r\s\Vert\l}\right)\uppsi^{(n-1)\,\cdots}{}_{\a\b\Vert}{}^\s,\eeq{xarb8}
which is our non-Abelian $2-s-s$ vertex with $2n-1$ derivatives. Comparing it with the spin-$\tf{5}{2}$
counterpart~\rf{3der8} reveals that they are very similar as well.

%%%%%%%%%%%%%%%%%%%%%%%%%%%%%%%%%%%%%%%%%%%%%%%%%%
%%%%%%%%%%%%%%%%%%%%%%%%%%%%%%%%%%%%%%%%%%%%%%%%%%
\subsection{Abelian Vertices}\label{sec:2theorems}
%%%%%%%%%%%%%%%%%%%%%%%%%%%%%%%%%%%%%%%%%%%%%%%%%%
%%%%%%%%%%%%%%%%%%%%%%%%%%%%%%%%%%%%%%%%%%%%%%%%%%

Abelian vertices are those that do not deform the gauge algebra. Such a vertex corresponds to a trivial $a_2$,
and therefore to an $a_1$ which can always be chosen to be $\G\text{-closed}$~\cite{BBH},
\beq \G a_1=0,\eeq{noa2strict}
and which is related to the vertex $a_0$ through the cocycle condition~(\ref{cocycle3}):
\beq \D a_1+\G a_0\doteq 0.\eeq{a0toa1local}

The $2-s-s$ Abelian vertices we wish to consider do not deform the gauge transformations either. In other words, for
gravitational cubic coupling of a higher-spin fermion the gauge symmetries remain intact unless the gauge algebra is
deformed.\footnote{The same is true for electromagnetic couplings of higher-spin fermions as well~\cite{EMours}. The
proof for the gravitational case is quite similar to that for the electromagnetic one.} To prove this, first we note
that it is always possible to rewrite a cubic vertex as \beq a_0=T^{\m\n}h_{\m\n},\eeq{a0currentdef} i.e., the graviton
field $h_{\m\n}$ contracted with a symmetric fermion-bilinear current $T^{\m\n}$. If the vertex is Abelian, we will see
that the latter can be chosen to satisfy \beq \G T^{\m\n}=0,\qquad \de_\n T^{\m\n}=\D M^\m\quad\text{with}\;\;\G M^\m=0.
\eeq{3cond} For $s=n+\tf{1}{2}$, let us write the most general form of the $a_1$ corresponding to~(\ref{a0currentdef}):
\beq a_1=2M^\m C_\m+\left(\bar{P}_{\m_1...\m_{n-1}}\xi^{\m_1...\m_{n-1}}-\bar{\xi}_{\m_1...\m_{n-1}}P^{\m_1...\m_{n-1}}
\right)+a_1',\eeq{a1froma0generalform} where $M^\m$ and $P_{\m_1...\m_{n-1}}$ belong to $H(\G)$ and have $pgh=0$, $agh=1$,
and $a_1'$ stands for expansion terms in the ghost-curls. The consistency condition~(\ref{a0toa1local}) now reads
\beq \G\left(T^{\m\n}h_{\m\n} \right)+2\D M^\m C_\m+\left(\D\bar{P}_{\m_1...\m_{n-1}}\xi^{\m_1...\m_{n-1}}
-\bar{\xi}_{\m_1...\m_{n-1}}\D P^{\m_1...\m_{n-1}}\right)+\D a_1'\doteq0.\eeq{a0toa1pluged}
It is clear from the properties of $P_{\m_1...\m_{n-1}}$ that it may consist of two kinds of terms: one contains the
antifield $h^{*\m\n}$ and its derivatives, and the other contains the antifield $\ps^{*\n_1...\n_n}$ and its derivatives.
The former kind also contains (derivatives of) the Fronsdal tensor $\mathcal S_{\n_1...\n_n}$ or (derivatives of) the
curvature $\Ps_{\m_1\n_1|...|\m_n\n_n}$, while the latter one contains (derivatives of) the linearized Riemann tensor
$R_{\m\n\r\s}$. By using the Leibniz rule, however, one can choose to get rid of derivatives on $h^{*\m\n}$ and $R_{\m\n\r\s}$.
Thus one can write
\bea P_{\m_1...\m_{n-1}}&=&h^{*\m\n}\left[\vec{P}^{(\mathcal S)}_{\m\n,\,\m_1...\m_{n-1}}{}^{\n_1...\n_n}\mathcal S_{\n_1...\n_n}+\vec{P}^{(\Ps)}_{\m\n,\,\m_1...\m_{n-1}}{}^{\n_1\r_1|...|\n_n\r_n}\Ps_{\n_1\r_1|...|\n_n\r_n}\right]
\nonumber\\&&+ R^{\m\n\r\s}\vec{P}^{(\ps^*)}_{\m\n\r\s,\,\m_1...\m_{n-1}}{}^{\n_1...\n_n}\ps^*_{\n_1...\n_n}
+\de^{\m_n}p_{\m_1...\m_n},\eea{Pmuexplicit}
where $\G p_{\m_1...\m_n}=0$, and the $\vec P$'s are differential operators acting to the right. Notice that in the above
expression both terms in the brackets are not only $\G$-closed but also $\D$-exact.\footnote{While the $\D$-exactness of the
first term therein is manifest, the second term contains the spin-$s$ curvature, which admits only $\D$-exact terms like its
own ($\g$-)traces and divergences~(see Appendix~\ref{sec:curvatures}), thanks to the way the indices are contracted.} Now,
taking the $\D$ variation of $P_{\m_1\dots\m_{n-1}}$ one finds from Eq.~\rf{Pmuexplicit} that
\beq \D P_{\m_1...\m_{n-1}}=\tf{1}{4}R^{\m\n\r\s}\D Q_{\m\n\r\s,\,\m_1...\m_{n-1}}+\de^{\m_n}\D q_{\m_1...\m_n},\eeq{DeltaPmu}
where the quantity $Q_{\m\n\r\s,\,\m_1...\m_{n-1}}$ is $\G$-closed and enjoys the same symmetries in its first four indices as
the Riemann tensor, and $\G q_{\m_1...\m_n}=0$. Therefore, one finds that
\beq \bar{\xi}^{\m_1...\m_{n-1}}\D P_{\m_1...\m_{n-1}}\doteq h^{\m\n}\D\left[\de^\a\de^\b\left(\bar\xi_{\m_1...\m_{n-1}}Q_{\m\a
\n\b,}{}^{\m_1...\m_{n-1}}\right)\right]-\bar{\xi}_{\m_1...\m_{n-1}}\overset\leftarrow\de_{\m_n}\D q^{\m_1...\m_n}.\eeq{xiPmu}
The last term on the right-hand side above is $\G$-closed, and can be broken into a $\G$-exact piece plus terms involving
the fermionic ghost-curls. The latter can always be canceled in the cocycle condition~(\ref{a0toa1pluged}) by appropriately
choosing $a_1'$. One is thus left with
\bea \G\left[T^{\m\n}h_{\m\n}+\D\left(\tf{1}{n}\bar\ps_{\m_1...\m_n}q^{\m_1\dots\m_n}+\text{h.c.}\right)\right]+2\D M^\m C_\m
~~~~~~~~~~~~~~~~~~~~~~~~~~\nonumber\\-h^{\m\n}\D\left[\de^\a\de^\b\left(\bar\xi_{\m_1...\m_{n-1}}Q_{\m\a\n\b,}{}^{\m_1...\m_{n-1}}
+\hc\right)\right]\doteq0.\eea{a0toa1new}
Now, one can drop the $\D$-exact terms added to the original vertex $T^{\m\n}h_{\m\n}$ to write \beq h^{\m\n}\left[\G T_{\m\n}
-\de^\a\de^\b\left(\bar\xi_{\m_1...\m_{n-1}}\D Q_{\m\a\n\b,}{}^{\m_1...\m_{n-1}}+\hc\right)\right]+2\left(\D M^\m
-\de_\n T^{\m\n}\right)C_\m\doteq0.\eeq{a0toa1newBIS} Taking a functional derivative w.r.t. $C_\m$ then yields the second condition
in Eq.~\rf{3cond}: \beq\de_\n T^{\m\n}=\D M^\m,\eeq{2ndcond} with $\G M^\m=0$ by assumption. On the other hand, a functional
derivative w.r.t. $h_{\m\n}$ gives
\beq \G T_{\m\n}=\de^\a\de^\b\left(\bar\xi_{\m_1...\m_{n-1}}\D Q_{\m\a\n\b,}{}^{\m_1...\m_{n-1}}\right)+\hc\,,\eeq{2ndfuncder}
which means, in particular, that the quantity on the right-hand side must be $\G$-exact. This is possible if
$\de^\a\de^\b Q_{\m\a\n\b,\,\m_1...\m_{n-1}}$ is $\D$-closed, and the indices of $Q$ have the interchange symmetry
$\a\leftrightarrow\m_i$ and $\b\leftrightarrow\m_i$ with $i=1,2,\dots,n-1$. This enables one to conclude
\beq T_{\m\n}=\tilde T_{\m\n}+\frac{1}{n}\,\D\left[2\bar\ps_{\m_1...\m_n}\de_\a Q_{(\m}{}^\a{}_{\n)}{}^{\m_1,\,\m_2...\m_n}
+\de_\a\bar\ps_{\m_1...\m_n} Q_{(\m}{}^\a{}_{\n)}{}^{\m_1,\,\m_2...\m_n}+\hc\right],\eeq{trivialT} where $\G\tilde T_{\m\n}=0$.
Therefore, one can render the current gauge invariant by field redefinitions without affecting the form~\rf{2ndcond} of its
divergence. This completes the proof of Eq.~\rf{3cond}. Then the $a_1$ following from Eq.~\rf{a0toa1local} reads
\beq a_1=2M^\m C_\m.\eeq{lifttoa1Abelian}

We will now prove a sufficient condition for the triviality of $a_1$, given by~\rf{lifttoa1Abelian}, and hence of the deformation
of the gauge transformations. It is
\beq \D M^\m=\de_\n\mathcal X^{(\m\n)}+\de_\r\de_\s\mathcal Y^{\m\r\s},\qquad \text{with }\mathcal X^{(\m\n)}, \mathcal Y^{\m\r\s}\;
\D\text{-exact and}~\G\text{-closed}.\eeq{abel0}
If Eq.~\rf{abel0} is true, then from Eq.~\rf{lifttoa1Abelian} we can write $\D a_1$ as
\beq \D a_1=2\left(\de_\n\mathcal X^{(\m\n)}+\de_\n\de_\r\mathcal Y^{\m\n\r}\right)C_\m\doteq-2\mathcal X^{(\m\n)}\de_{(\m}
C_{\n)}+2\mathcal Y^{\m\n\r}\de_\n\de_\r C_\m.\eeq{abel1}
But the derivatives of the bosonic ghost are $\G$-exact: $2\de_{(\m}C_{\n)}=\G h_{\m\n}$ and $2\de_\n\de_\r C_\m=\de_\r\G h_{\m\n}
-\G\mathfrak h_{\m\n\Vert\r}$. Because $\mathcal X^{(\m\n)}$ and $\mathcal Y^{\m\n\r}$ are $\G$-closed, one can write
\beq \D a_1\doteq-\G\left[\left(\mathcal X^{(\m\n)}+\de_\r\mathcal Y^{\m\n\r}\right)h_{\m\n}+\mathcal Y^{\m\n\r}
\mathfrak h_{\m\n\Vert\r}\right].\eeq{abel2}
In view of the the cocycle condition $\D a_1\doteq-\G a_0$, one can therefore write
\beq \G\left[a_0-\left(\mathcal X^{(\m\n)}+\de_\r\mathcal Y^{\m\n\r}\right)h_{\m\n}-\mathcal Y^{\m\n\r}\mathfrak h_{\m\n\Vert\r}
\right]\doteq0.\eeq{abel3}
Because the quantities added to $a_0$ on the left-hand side are $\D$-exact by assumption, one can render the vertex
gauge-invariant only up to a total derivative, by field redefinitions. This proves the triviality of $a_1$ if Eq.~\rf{abel0}
holds.

The arguments presented in the beginning of Subsection~\ref{subsec:Abelian} go verbatim for arbitrary spin, except that now
the number of derivatives in the Abelian vertex can take the values $2n, 2n+1$ and $2n+2$, since the spin-$s$ curvature
tensor contains $n$ derivatives. The corresponding currents can be written as direct generalizations of those for spin $\tf{5}{2}$,
given respectively by Eq.~\rf{4der0} with $\a=\tf{1}{4}$, and by Eqs.~\rf{5der0} and~\rf{6der0}. The explicit vertices are:
\bea &p=2n:~~a_0=ig\left(h_{\m\n}-\tf{1}{4}\h_{\m\n}h'\right)\bar\Ps^\m{}_{\cdots}\Ps^{\n\,\cdots}\,,&\label{abelderp0}\\
&p=2n+1:~~a_0=ig\,h_{\m\n}\bar\Ps^{\,\cdots}\,\g^{(\m}\overset\leftrightarrow\de{}^{\n)}\Ps_{\cdots}\,,&\label{abelderp1}\\
&p=2n+2:~~a_0=ig\,h_{\m\n}\bar\Ps^{\,\cdots}\left(\overset\rightarrow{\de}{}^\m\overset\rightarrow{\de}{}^\n
+\overset\leftarrow{\de}{}^\m\overset\leftarrow{\de}{}^\n-\h^{\m\n}\overset\leftarrow{\de}{}^\l\overset\rightarrow{\de}
{}_\l\right)\Ps_{\cdots}\,.&\eea{abelderp2}

None of these vertices deform the gauge transformations. The $2n$-derivative vertex can be shown to fulfill the sufficient
condition~\rf{abel0} in order for its $a_1$ to be trivial, and the proof follows exactly the same steps as in the spin-$\tf{5}{2}$ case.
On the other hand, one can render the $(2n+1)$-derivative vertex manifestly $\G$-closed modulo $d$ by casting it into a generalization
of Eq.~\rf{5der2}, while the $(2n+2)$-derivative one takes the 3-curvature form like Eq.~\rf{6der1}. These proofs are also
straightforward generalizations of the spin-$\tf{5}{2}$ case.

Finally, direct generalizations of the prototypical spin-$\tf{5}{2}$ example also show that the $2n$- and $(2n+1)$-derivative
vertices are trivial in $D=4$, while the $(2n+2)$-derivative 3-curvature vertex exits in all $D\geq 4$.

%%%%%%%%%%%%%%%%%%%%%%%%%%%%%%%%%%%%%%%%%%%%%%%%
%%%%%%%%%%%%%%%%%%%%%%%%%%%%%%%%%%%%%%%%%%%%%%%%
\section{Beyond Cubic Order}\label{sec:2ndorder}
%%%%%%%%%%%%%%%%%%%%%%%%%%%%%%%%%%%%%%%%%%%%%%%%
%%%%%%%%%%%%%%%%%%%%%%%%%%%%%%%%%%%%%%%%%%%%%%%%

Let us recall from Eq.~\rf{brst2nd} that consistent second-order deformations require
\beq (S_1,S_1)=-2sS_2=-2\Gamma S_2-2\Delta S_2.\eeq{2nd1} This antibracket is zero for the Abelian vertices, which go unobstructed
beyond the cubic level. The non-Abelian vertices, on the other hand, have nontrivial $a_1$ and $a_2$ and may not fulfill this
requirement. Here we will prove by contradiction that indeed they do not.

Notice that $S_2$ is at most linear in the antifields $\Phi^*_A$, on which $\G$ does not act. On the other hand, the $\D$ variation
of only an antighost $\mathcal C^*_\a$ may produce an antifield. Therefore, the general form of the antibracket evaluated at zero
antifields is
\beq \left[(S_1,S_1)\right]_{\Phi^*_A=0}=\G N+\D M,\eeq{2nd2}
with $N\equiv -2\left[S_2\right]_{\Phi^*_A=0}$ and $M\equiv -2\left[S_2\right]_{\mathcal C^*_\a=0}$.
Let us also note that in the antibracket of $S_1=\int(a_2+a_1+a_0)$ with itself, among all the possibilities, only the antibracket
between $\int a_0$ and $\int a_1$ survives when the antifields are set to zero.  Thus one is left with
\beq \left[(S_1,S_1)\right]_{\Phi^*_A=0}=2\left(\int a_0,\int a_1\right)\equiv\int b.\eeq{2nd6}
It is relatively easier to compute the quantity $b$, which must satisfy the requirement:
\beq b\doteq\G\text{-exact}+\D\text{-exact},\eeq{2nd7}
in view of Eqs.~\rf{2nd2} and~\rf{2nd6}. For simplicity, we stick to the prototypical spin-$\tf{5}{2}$ case.

%%%%%%%%%%%%%%%%%%%%%%%%%%%%%%%%%%%%%%%%%
\subsubsection*{The 2-Derivatives Vertex}
%%%%%%%%%%%%%%%%%%%%%%%%%%%%%%%%%%%%%%%%%

For the 2-derivative $2-\tfrac{5}{2}-\tfrac{5}{2}$ vertex, let us write down the deformations $a_0$ and $a_1$. First, from
Eq.~\rf{2dervertex1}, one can rewrite the vertex as $a_0\doteq T^{\m\n}h_{\m\n}$. The result is
\beq T^\m{}_\n=4ig\left[\de^\r\de_\s\left\{\bar\ps_{\n\l}\left(\h^{\m\s|\l\t}+\tf{1}{2}\g^{\m\s\l\t}\right)\ps_{\r\t}
+\tf{1}{2}\bar{\pss}_{[\n}\g^{\m\s}\pss_{\r]}\right\}+\tf{1}{16}\bar\uppsi_{\r\s\Vert}{}^\l\,\g^{\m\r\s\a\b,}{}_{\n\l\g}
\,\uppsi_{\a\b\Vert}{}^\g\right].\eeq{quar2a0}
On the other hand, from Eqs.~\rf{2da1}--\rf{a1gt} and Eq.~\rf{2deramb}, we can write
\beq a_1=ig\,h^{*\m}{}_\n\left(\bar\upxi_{\m\l}\ps^{\n\l}+\bar\ps^{\n\l}\upxi_{\m\l}-2\bar\xi^\l\uppsi_{\m\l\Vert}{}^\n
-2\bar\uppsi_{\m\l\Vert}{}^\n\xi^\l+\G j_\m{}^\n\right)+\cdots,\eeq{quar2a1}
where the ellipses stand for terms containing the antifield $\ps^{*\m\n}$, and $j_\m{}^\n$ is some spin-$\tf{5}{2}$ bilinear.
Then the quantity $b$ will contain 4-fermion terms plus fermion bilinears:\footnote{The latter terms, which we will not
write explicitly, come from the ellipses in Eq.~\rf{quar2a1}.}
\beq b=2ig\,T^{\m}{}_\n\left(\bar\upxi_{\m\l}\ps^{\n\l}+\bar\ps^{\n\l}\upxi_{\m\l}
-2\bar\xi^\l\uppsi_{\m\l\Vert}{}^\n-2\bar\uppsi_{\m\l\Vert}{}^\n\xi^\l+\G j_\m{}^\n\right)+\cdots\,.\eeq{quar21}
Note that the two kind of terms are completely different and we can treat them separately. If Eq.~\rf{2nd7}
is fulfilled, a functional derivative thereof w.r.t.~$\bar\xi_\m$ has to be $\D$-exact plus the divergence of a
symmetric tensor. This functional derivative reads:
\beq \frac{\d b}{\d\bar{\xi}_\m}=4ig\left[\de_\n\left(T^{[\m}_{\r}\ps^{\n]\r}\right)+T_{\r\s}\uppsi^{\m\r\Vert\s}
\right]+\cdots\,.\eeq{quar22}
Because the vertex is nontrivial, $T^{\m}{}_\n$ cannot be $\D$-exact. Now, the right-hand side of Eq.~\rf{quar22}
is trilinear in the spin-$\tf{5}{2}$ field. Given that possible Fierz rearrangements cannot redistribute the derivatives
among the fields, let us consider, among others, the terms in which three derivatives act on a single fermion.
By inspection, it is clear that these terms cannot be written as  $\D$-exact plus the divergence of a symmetric tensor.

Therefore, it is not possible to satisfy Eq.~\rf{2nd7}. Then in a local theory the non-Abelian $2-\tfrac{5}{2}-\tfrac{5}{2}$
vertex with two derivatives gets obstructed beyond the cubic order.

%%%%%%%%%%%%%%%%%%%%%%%%%%%%%%%%%%%%%%%%%
\subsubsection*{The 3-Derivatives Vertex}
%%%%%%%%%%%%%%%%%%%%%%%%%%%%%%%%%%%%%%%%%

The proof for the 3-derivative case is in the same spirit as the previous example. Let us rewrite, from
Eq.~\rf{3der8}, the vertex as $a_0\doteq T^{\m\n}h_{\m\n}$, with the current given by
\beq T_\m{}^\n=2ig\de^\l\left(\bar{\uppsi}_{\r\s\Vert\m}\,\g^{\n\r\s\a\b}\,\uppsi_{\a\b\Vert\l}
-\bar{\uppsi}_{\r\s\Vert\l}\,\g^{\n\r\s\a\b}\,\uppsi_{\a\b\Vert\m}+\bar{\uppsi}_{\r\s\Vert}{}^\g\,
\g^{\n\r\s\a\b,}{}_{\m\l\g\d}\,\uppsi_{\a\b\Vert}{}^\d\right).\eeq{quar3a0}
Now $a_1$ is given by Eqs.~\rf{3der2} and~\rf{3der6}; it has the form:
\beq a_1=-2igh_\m^{*\,\n}\left(\bar{\upxi}_{\r\s}\g^{\m\r\s\a\b}\uppsi_{\a\b\Vert\n}-\bar\uppsi_{\r\s\Vert\n}\g^{\m\r\s\a\b}
\upxi_{\a\b}\right)+\cdots .\eeq{quar3a1}
Again, the quantity $b$ will contain 4-fermion terms and fermion bilinears:
\beq b=-4ig\,T_\m{}^\n\left(\bar{\upxi}_{\r\s}\g^{\m\r\s\a\b}\uppsi_{\a\b\Vert\n}
-\bar\uppsi_{\r\s\Vert\n}\g^{\m\r\s\a\b}\upxi_{\a\b}\right)+\cdots .\eeq{quar31}
Again, let us consider, in the functional derivative of $b$ w.r.t.~$\bar\xi_\m$, the terms with three derivatives acting
on a single fermion to find that they cannot be written as  $\D$-exact plus the divergence of a symmetric tensor.
Therefore, Eq.~\rf{2nd7} will not be satisfied, and so in a local theory,
the $3$-derivative $2-\tfrac{5}{2}-\tfrac{5}{2}$ vertex is also inconsistent at the quartic order.

Let us note that there is no such obstruction for the cubic gravitational couplings of a spin-$\tf{3}{2}$ field,
and indeed one can find a local theory consistent to all orders: $\mathcal N=1$ supergravity~\cite{N=1SUGRA}. For
$s\geq\tf{5}{2}$, the obstruction will hold even if one considers an arbitrary linear combination of the non-Abelian and
Abelian vertices.
It is expected, however, that in a theory consistent beyond the cubic level only a specific linear combination of the
vertices survives. This is indeed the case with the tensionless limit of open string theory~\cite{Taronna}. We emphasize
that, for higher spins, the obstruction is removed if one gives up locality. The call for non-locality may be intrinsic
or may result from having integrated out additional dynamical fields present in the consistent interacting theory or both.
String theory, for example, realizes the higher-order consistency by invoking them both~\cite{4-point}.

%%%%%%%%%%%%%%%%%%%%%%%%%%%%%%%%%%%%%%%%%%%%%%%
%%%%%%%%%%%%%%%%%%%%%%%%%%%%%%%%%%%%%%%%%%%%%%%
\section{Concluding Remarks}\label{sec:remarks}
%%%%%%%%%%%%%%%%%%%%%%%%%%%%%%%%%%%%%%%%%%%%%%%
%%%%%%%%%%%%%%%%%%%%%%%%%%%%%%%%%%%%%%%%%%%%%%%

In this paper, we have constructed parity-preserving covariant cubic vertices for arbitrary-spin fermionic gauge fields
coupled to gravity in flat space, by employing the BRST-BV cohomological methods.\footnote{As has been emphasized
in~\cite{Metsaev-BRST}, the BRST-BV approach is very useful  in general for obtaining gauge-invariant manifestly
Lorentz-invariant off-shell vertices for higher-spin fields~\cite{Tsulaia}.} We have seen that gauge invariance and
non-triviality of the deformations rule out minimal coupling, and constrain the number of derivatives in a $2-s-s$
vertex, in complete accordance with what the light-cone formulation reveals~\cite{Metsaev}. Two of the lowest-derivative
vertices call for deformations of the gauge algebra, i.e., they are non-Abelian. They cannot be extended in a local theory
beyond the cubic order in the absence of additional interacting dynamical fields. It turns out that none of the
gauge-algebra-preserving Abelian vertices deform the gauge transformations.

Our covariant off-shell cubic vertices must be equivalent to those inspired by open string theory, reported in~\cite{Taronna},
for an obvious reason: both results are in one-to-one correspondence with the light-cone-formulation ones, and for each
allowed derivative value there is a unique vertex~\cite{Metsaev_Boson,Metsaev}. We do not present a direct demonstration
of this equivalence, which, however, was done for the electromagnetic case~\cite{EMours}. As we already mentioned, the
string-inspired vertices all come with a fixed known numerical coefficient, with the single dimensionful coupling constant
set to unity. The apparent freedom for each of our coupling constants may very well be an artifact of the cubic-order
analysis that disappears once higher-order consistency is taken into account~\cite{BBH}. The more rigid structure in the
string-inspired interactions may not be surprising then, since string theory is consistent beyond the cubic order, with
non-locality necessarily creeping in~\cite{4-point}.

Let us see how the $2-s-s$ vertices differ for fermionic and bosonic higher spins. First, a boson possesses only three
such vertices: with $2s-2$, $2s$ and $2s+2$ derivatives, the jump in the number of derivatives being
two~\cite{2-3-3,2-s-s,Metsaev_Boson}. In contrast, we have seen that a fermion has five cubic vertices; the derivatives
range from $2s-3$ to $2s+1$, with a jump of unity. In both cases, though, the number of nontrivial vertices in 4D
is the same: two. In $D\geq5$, a fermion exceeds a boson by two vertices: one non-Abelian, another Abelian.
On the other hand, $1-s-s$ electromagnetic couplings for fermions also have jump of unity in the number of derivatives,
which may take three values: $2s-2$, $2s-1$ and $2s$. Again, two vertices with the lowest and highest number of derivatives
survive in 4D.

Minimal gravitational coupling of massless higher spins does exist in AdS space~\cite{FV}. The Fradkin-Vasiliev
construction~\cite{FV} incorporates only the cubic non-Abelian vertices. Although the interactions are non-analytic
in the cosmological constant, one can take the $\L\rightarrow0$ limit judiciously to be left with the highest-derivative
non-Abelian vertex~\cite{2-s-s}. It is expected that our flat-space non-Abelian vertices are present in the
Fradkin-Vasiliev system in AdS, and a careful flat limit should pick them up. The study of gravitational interaction
vertices of a massive spin-$\tf{5}{2}$ field in AdS was actually carried out in~\cite{Metsaev:2006ui}, where it was
noticed that what survives in the massless flat limit is only a 2-derivative vertex when $D=4$, or a 3-derivative one
when $D>4$. This must precisely be our flat-space highest-derivative non-Abelian cubic vertex in the respective dimensions.

What connection may our vertices have with the massive theory? For a massive spin-$\tf{5}{2}$ field, coupled to
gravity in flat space, it was noticed in Ref.~\cite{Porrati:1993in} that suitable non-minimal couplings improve the high-energy
behavior of the theory by pushing higher the scale at which tree-level unitarity is violated. The simplest of these
terms has two derivatives: in 4D it reads $\bar\ps_{\m\a}R^{+\m\n\a\b}\ps_{\n\b}$ modulo on-shell terms. This is nothing
but the first piece in our 2-derivative vertex~\rf{2dervertex1}$-$the part surviving in 4D in the transeverse-traceless
gauge. This may not come as a surprise. After all, consistent massive theories are expected to originate from massless ones.
A similar thing happens for the spin-$\tf{3}{2}$ electromagnetic coupling~\cite{EMours}: the gauge-invariant
Pauli term $\bar\ps_\m F^{+\m\n}\ps_\n$ does improve the tree-level unitarity of the massive theory~\cite{Ferrara:1992yc},
and shows up in the consistent $\mathcal N=2$ broken supergravity theory~\cite{BrokenSUGRA}.

It would be interesting to extend our systematic analysis to (A)dS space.
There are certain technical difficulties, though, in extending the applicability of the
BRST deformation scheme to spaces of constant curvature. One may use the ambient-space formulation~\cite{Ambient} for
AdS space, in particular, to avoid these issues. Then one could construct covariant higher-spin vertices in AdS, and
the results could be compared with those obtained recently in Refs.~\cite{Cubic_AdS0,Cubic_AdS}. This would help
us understand better the rather intricate structure of the Vasiliev higher-spin systems~\cite{Vasiliev}, by possibly
leading us a step closer to a yet-to-be-found standard action. We leave constructions in AdS space as future work.

%%%%%%%%%%%%%%%%%%%%%%%%%%%%%
\subsection*{Acknowledgments}
%%%%%%%%%%%%%%%%%%%%%%%%%%%%%

We would like to thank P.~Benincasa, N.~Boulanger, A.~Campoleoni, E.~Conde Pena, R.~Metsaev, M.~Taronna and M.~Vasiliev for
useful discussions. MH gratefully acknowledges support from the Alexander von Humboldt Foundation through a Humboldt Research
Award and support from the ERC through the ``SyDuGraM" Advanced Grant. GLG is a Research Fellow of the Fonds pour la Formation
\`a la Recherche dans l'Industrie et dans l'Agriculture (F.R.I.A.), who is partially supported by IISN-Belgium
(convention 4.4514.08). RR is a Postdoctoral Fellow of the Fonds de la Recherche Scientifique-FNRS, whose work is
partially supported by IISN-Belgium (conventions 4.4511.06 and 4.4514.08). Both GLG and RR are also partially supported
by the ``Communaut\'e Fran\c{c}aise de Belgique" through the ARC program and by the ERC Advanced Grant ``SyDuGraM."

\newpage
\begin{appendix}
\numberwithin{equation}{section}

%%%%%%%%%%%%%%%%%%%%%%%%%%%%%%%%%%%%%%%%%%%%%%%%%%%%%%%%%%%%%%%%
%%%%%%%%%%%%%%%%%%%%%%%%%%%%%%%%%%%%%%%%%%%%%%%%%%%%%%%%%%%%%%%%
\section{Curvatures, Identities \& EoMs}\label{sec:curvatures}
%%%%%%%%%%%%%%%%%%%%%%%%%%%%%%%%%%%%%%%%%%%%%%%%%%%%%%%%%%%%%%%%
%%%%%%%%%%%%%%%%%%%%%%%%%%%%%%%%%%%%%%%%%%%%%%%%%%%%%%%%%%%%%%%%

In this Appendix we will discuss some important properties of the curvatures and curls of the different
fields under consideration. We will also write down various forms of the EoMs in terms of these objects,
which would help us identify $\D$-exact terms.

%%%%%%%%%%%%%%%%%%%%
\subsection*{Spin 2}
%%%%%%%%%%%%%%%%%%%%

The 1-curl, $\mathfrak{h}_{\m\n\Vert\r}$, of the spin-2 field is antisymmetric in its first two indices, and obeys
the Bianchi identities, \beq \mathfrak{h}_{[\m\n\Vert\r]}=0\quad\Leftrightarrow\quad\mathfrak{h}_{\m[\n\Vert\r]}
=-\tf{1}{2}\mathfrak{h}_{\n\r\Vert\m},\qquad \de_{[\m}\mathfrak{h}_{\n\r]\Vert\s}=0.\eeq{x1} The linearized Riemann
tensor, $R_{\m\n\r\s}=R_{[\m\n][\r\s]}=R_{\r\s\m\n}$, obeys the same: \beq R_{[\m\n\r]\s}=0,\qquad \de_{[\m}R_{\n\r]
\a\b}=0.\eeq{x2}
The original graviton EoMs are given in terms of the linearized Einstein tensor,
\beq G_{\m\n}= R_{\m\n}-\tf{1}{2}\h_{\m\n}R=\D h^*_{\m\n}.\eeq{grav1}
Taking a trace, it follows immediately that \beq R_{\m\n}=\D\left(h^*_{\m\n}-\tf{1}{D-2}\h_{\m\n}h^{*\prime}\right),
\qquad R=-\left(\tf{2}{D-2}\right)\D h^{*\prime},\eeq{x4} where ``prime'' denotes a trace, $h^{*\prime} \equiv h^{*\m}_{~\m}$.
The second one in Eq.~\rf{x2} leads to the contracted Bianchi identity, which says that the divergence of the Riemann
tensor is $\D$-exact:
\beq \de^\r R_{\m\n\r\s}=2\de_{[\m}R_{\n]\s}=2\de_{[\m}G_{\n]\s}-\h_{\s[\m}\de_{\n]}R=\D\text{-exact}.\eeq{grav2}
A trace of the above identity shows that the Einstein tensor is divergenceless, $\de^\m G_{\m\n}=0$, while a
double $\g$-trace gives
\beq \de^\r\Rs_{\r\s}=\left(\g^\m\g^\n-\h^{\m\n}\right)\left(2\de_{\m}G_{\n\s}-\h_{\s\m}\de_\n R\right)=
2\ds\Gs_\s-\g_\s\ds R+\ds_\s R=\D\text{-exact}.\eeq{grav3}
By using the relation $\g^\a\g^{\r\s}=2\g^{\a\r\s}-\g^{\r\s}\g^\a$, and the first Bianchi identity in Eq.~\rf{x2},
it is easy to see that $\ds\,\Rs_{\m\n}=-\Rs_{\m\n}\overleftarrow{\ds}$. This quantity is actually $\D$-exact:
\beq \ds\,\Rs_{\m\n}=-\Rs_{\m\n}\overleftarrow{\ds}=4\g^\r\de_{[\m}R_{\n]\r}=\D\text{-exact}.\eeq{x3}
Other forms of $\D$-exact terms, that we do not use, include $\g^\m\Rs_{\m\n}$ and $\Box R_{\m\n\r\s}$.

%%%%%%%%%%%%%%%%%%%%%%%%%%%%%%%%%%%%%%%%%%%%%%%%%
\subsection*{Spin \textbf{5/2}}\label{subsec:5/2}
%%%%%%%%%%%%%%%%%%%%%%%%%%%%%%%%%%%%%%%%%%%%%%%%%

Just like spin 2, the 1-curl of the spin-$\tf{5}{2}$ field, $\uppsi_{\m\n\Vert\r}$, is antisymmetric in its first two
indices, and obeys the Bianchi identities, \beq \uppsi_{[\m\n\Vert\r]}=0\quad\Leftrightarrow\quad\uppsi_{\m[\n\Vert\r]}
=-\tf{1}{2}\uppsi_{\n\r\Vert\m},\qquad \de_{[\m}\uppsi_{\n\r]\Vert\s}=0.\eeq{x11} The curvature tensor, $\Ps_{\m\n|\r\s}
=\Ps_{[\m\n]|[\r\s]}=\Ps_{\r\s|\m\n}$, obeys the same: \beq \Ps_{[\m\n|\r]\s}=0,\qquad \de_{[\m}\Ps_{\n\r]|\a\b}=0.\eeq{x12}
For spin $\tf{5}{2}$, let us recall from Section~\ref{sec:5/2}, that the original EoMs are given by
\sea{nonmame2}{\mathcal R_{\m\n}&=\mathcal{S}_{\m\n}-\g_{(\m}\displaystyle{\not{\!\mathcal{S}}}_{\n)}-\tf{1}{2}
\h_{\m\n}\mathcal{S}'=\D \ps^{*}_{\m\n},\label{eom5/2}\\\bar{\mathcal R}_{\m\n}&=\bar{\mathcal S}_{\m\n}
-\displaystyle{\not{\!\bar{\mathcal S}}}_{(\m}\g_{\n)}-\tf{1}{2}\h_{\m\n}\bar{\mathcal S}'
=\D \bar{\ps}^{*}_{\m\n}.\label{eom5/2bar}}
One can easily rewrite these in terms of the Fronsdal tensor,
\beq \mathcal S_{\n_1\n_2}\equiv i\left[\ds\,\ps_{\n_1\n_2}-2\de_{(\nu_1}\pss_{\nu_2)}\right]=\D\vf^*_{\n_1\n_2},
\eeq{eom5/21}
and similarly $\bar{\mathcal{S}}_{\n_1\n_2}=\D\bar{\vf}^*_{\n_1\n_2}$ for its Dirac conjugate, where
\beq \vf^*_{\m\n}\equiv\ps^*_{\m\n}-\tf{2}{D}\g_{(\m}\pss^*_{\n)}-\tf{1}{D}\h_{\m\n}\ps^{*\prime}.\eeq{x10}
From the definition of the Fronsdal tensor, one easily finds that \beq \g^\s\uppsi_{\r\s\Vert\a}=i\mathcal{S}_{\r\a}
-\de_\a\pss_\r,\eeq{x13} whose $\g$-trace, in turn, gives: \beq \g^{\r\s}\uppsi_{\r\s\Vert\a}
=i\displaystyle{\not{\!\mathcal{S}}}_\a-\de_\a\ps',\qquad \psi'=\ps^\m_\m.\eeq{x14}

Now we see that the quantity $\g^{\m_1}\Ps_{\m_1\n_1|\,\m_2\n_2}$ is given by the $1$-curl of Eq.~\rf{x13},
and that it is $\D$-exact:
\beq \g^{\m_1}\Ps_{\m_1\n_1|\,\m_2\n_2}=-2i\de_{[\mu_2}\mathcal{S}_{\n_2]\n_1}=\D\text{-exact}.\eeq{Fronsdal1curl}
Similarly, from a 1-curl of Eq.~\rf{x14}, we obtain another useful form:
\beq \g^{\m_1\n_1}\Ps_{\m_1\n_1|\,\m_2\n_2}=2i\g^{\n_1}\de_{[\mu_2}\mathcal{S}_{\n_2]\nu_1}=\D\text{-exact}.
\eeq{eom5/24}
Taking a curl of~\rf{Fronsdal1curl}, one finds yet another form,
\beq \ds\,\Ps^{\m_1\n_1|}{}_{\m_2\n_2}=-4i\de^{[\mu_1}\de_{[\mu_2}\mathcal{S}_{\n_2]}{}^{\nu_1]}=\D\text{-exact}.
\eeq{eom5/25}
Eqs.~\rf{Fronsdal1curl}--\rf{eom5/25} also mean $\Ps^\m{}_{\n|\m\s}=\D\text{-exact}$ and
$\Box\Ps_{\m\n|\r\s}=\D\text{-exact}$. Finally, by using the identity $\de^{\m_1}=\tf{1}{2}\left(\ds\g^{\m_1}
+\g^{\m_1}\ds\right)$, we derive from Eqs.~\rf{Fronsdal1curl} and~\rf{eom5/25} that
\beq \de^{\m_1}\Ps_{\m_1\n_1|\,\m_2\n_2}=-2i\ds\,\de_{[\mu_2}\mathcal{S}_{\n_2]\n_1}+i\de_{\n_1}\g^\r
\de_{[\mu_2}\mathcal{S}_{\n_2]\r}=\D\text{-exact}.\eeq{eom5/26}
Similarly, one can find the various forms of the EoMs for the Dirac conjugate spinor.

Now from the definition of the Fronsdal tensor, one can find the identity
\beq \de\cdot\mathcal{S}_\m=\tf{1}{2}\ds\,\displaystyle{\not{\!\mathcal{S}}}_\m+\tf{1}{2}\de_\m\mathcal{S}'.
\eeq{x17}
Taking a divergence of Eq.~\rf{eom5/2}, and then using the above identity, one can write
\beq \de_\n\mathcal{R}^{\m\n}=-\tfrac{1}{2}\g^\m\,\de\cdot{\displaystyle{\not{\!\mathcal{S}}}}.\eeq{Noether-21}
This can be rewritten, by using Eqs.~\rf{nonmame2}--\rf{x10}, as
\beq \D\left(\de_\n\chi^{*\m\n}\right)=0,\qquad\chi^{*\m\n}\equiv\ps^{*\m\n}-\tf{1}{D}\,\g^\m\pss^{*\n}.\eeq{N23}
%

%%%%%%%%%%%%%%%%%%%%%%%%%%%%%%%%%%%%%%%%%%%%%%%%%%%%
\subsection*{Arbitrary Spin}\label{subsec:arbitrary}
%%%%%%%%%%%%%%%%%%%%%%%%%%%%%%%%%%%%%%%%%%%%%%%%%%%%

Let us recall that for arbitrary spin $s=n+\tf{1}{2}$, we have a totally symmetric rank-$n$ tensor-spinor
$\ps_{\n_1...\n_n}$, whose curvature is its $n$-curl, i.e., the rank-$2n$ tensor
\beq \Ps_{\m_1\n_1|\m_2\n_2|...|\m_n\n_n}=\left[...\left[\,\left[\de_{\m_1}...\de_{\m_n}\ps_{\n_1...\n_n}-
(\m_1\leftrightarrow\n_1)\right]-(\m_2\leftrightarrow\n_2)\right]...\right]-(\m_n\leftrightarrow\n_n).\eeq{curv1}
The curvature tensor~(\ref{curv1}) is gauge invariant even for an unconstrained gauge parameter. Its properties
can be found in~\cite{Curvature}. The curvature is antisymmetric under the interchange of ``paired'' indices, e.g.,
\beq \Ps_{\m_1\n_1|\m_2\n_2|\dots|\m_n\n_n}=-\Ps_{\n_1\m_1|\m_2\n_2|\dots|\m_n\n_n},\eeq{curv2}
but symmetric under the interchange of any two sets of paired indices, e.g.,
\beq \Ps_{\m_1\n_1|\m_2\n_2|\dots|\m_{n-1}\n_{n-1}|\m_n\n_n}=\Ps_{\m_n\n_n|\m_2\n_2|\dots|\m_{n-1}\n_{n-1}|\m_1\n_1}.
\eeq{curv3}
Another important property of the curvature is that it obeys the Bianchi identities
\beq \Ps_{[\m_1\n_1|\m_2]\n_2|...|\m_n\n_n}=0,\qquad \de_{[\r}\Ps_{\m_1\n_1]|\m_2\n_2|...|\m_n\n_n}=0.\eeq{curv4BIS}
Actually, these properties hold good for any $m$-curl, $m\leq n$, that contains paired indices.

For spin $s=n+\tf{1}{2}$, we recall from Section~\ref{sec:arbitrary} that the original EoMs read
\sea{noname4}{\mathcal{R}_{\m_1 \dots \m_n} &=\mathcal S_{\m_1...\m_n}-\tf{1}{2}n\,\g_{(\m_1}
\displaystyle{\not{\!\mathcal S}}_{\m_2...\m_n)}-\tf{1}{4}n(n-1)\,\h_{(\m_1\m_2}\mathcal S^\prime_{\m_3
...\m_n)}=\D \ps^{*}_{\m_1 \dots \m_n},\label{eomarb}\\\bar{\mathcal{R}}_{\m_1 \dots \m_n}&= \mathcal
{\bar S}_{\m_1...\m_n}-\tf{1}{2}n\displaystyle{\not{\!\mathcal{\bar S}}}_{(\m_1...\m_{n-1}}\g_{\m_n)}
-\tf{1}{4}n(n-1)\,\h_{(\m_1\m_2}\mathcal{\bar S}^\prime_{\m_3...\m_n)}=\D \bar{\ps}^{*}_{\m_1...
\m_n}.\label{eomarbbar}}
One can reexpress the EoMs in terms of the Fronsdal tensor as follows: \beq \mathcal{S}_{\n_1...\n_n}
\equiv i\left[\ds\,\ps_{\n_1...\n_n}-n\de_{(\n_1}\pss_{\n_2...\n_n)}\right]=\D\vf^*_{\n_1...\n_n},\eeq{eomarb1}
and similarly $\bar{\mathcal{S}}_{\n_1...\n_n}=\D\bar{\vf}^*_{\n_1...\n_n}$ for its Dirac conjugate, where
\beq \vf^*_{\n_1...\n_n}\equiv \ps^*_{\n_1...\n_n}-\tf{n}{2n+D-4}\,\g_{(\n_1}\pss^*_{\n_2...\n_n)}-\tf{n(n-1)}
{2(2n+D-4)}\,\h_{(\n_1\n_2}\ps^{*\prime}_{\n_3...\n_n)}.\eeq{x20}
Taking an $(n-2)$-curl of the the Fronsdal tensor~\rf{eomarb1}, one finds the relation
\beq \g^{\n_{n-1}}\uppsi^{(n-1)}_{\m_1\n_1|...|\,\m_{n-1}\n_{n-1}\Vert\n_n}=i\mathcal{S}^{(n-2)}_{\m_1\n_1|
...|\,\m_{n-2}\n_{n-2}\Vert\n_{n-1}\n_n}-\de_{\n_n}\displaystyle{\not{\!\uppsi}}^{(n-2)}_{\m_1\n_1|...|\,
\m_{n-2}\n_{n-2}\Vert\n_{n-1}},\eeq{x23} whose $\g$-trace, in turn, gives:
\beq \g^{\m_{n-1}\n_{n-1}}\uppsi^{(n-1)}_{\m_1\n_1|...|\,\m_{n-1}\n_{n-1}\Vert\n_n}=i\displaystyle{\not
{\mathcal{\!S}}}^{(n-2)}_{\m_1\n_1|...|\,\m_{n-2}\n_{n-2}\Vert\n_n}-\de_{\n_n}\uppsi^{\prime(n-2)}
_{\m_1\n_1|...|\,\m_{n-2}\n_{n-2}}.\eeq{x24}
The arbitrary spin generalizations of Eqs.~\rf{Fronsdal1curl}--\rf{eom5/26} are rather straightforward,
and can be derived the same way. They respectively read
\bea &\g^{\m_1}\Ps_{\m_1\n_1|...|\,\m_n\n_n}=-i\mathcal{S}^{(n-1)}_{\m_2\n_2|...|\m_n\n_n\Vert\n_1}=\D
\text{-exact},&\label{eomarb2}\\&\g^{\m_1\n_1}\Ps_{\m_1\n_1|...|\,\m_n\n_n}=i\displaystyle{\not{\!\mathcal S}}
^{(n-1)}_{\m_2\n_2|...|\,\m_n\n_n}=\D\text{-exact},&\label{eomarb3}\\&\ds\,\Ps_{\m_1\n_1|...|\,\m_n\n_n }
=-i\mathcal{S}^{(n)}_{\m_1\n_1|...|\,\m_n\n_n}=\D\text{-exact},&\label{eomarb4}\\&\de^{\m_1}\Ps_{\m_1\n_1|...|
\,\m_n\n_n}=-i\ds\,\mathcal{S}^{(n-1)}_{\m_2\n_2|...|\m_n\n_n\Vert\n_1}+\tf{i}{2}\de_{\n_1}
\displaystyle{\not{\!\mathcal{S}}}^{(n-1)}_{\m_2\n_2|...|\m_n\n_n}=\D\text{-exact}.&\eea{eomarb5}
Obvious consequences of the above equations include the $\D$-exactness of $\h^{\m_1\m_2}\Ps_{\m_1\n_1|...|\,\m_n\n_n}$
and $\Box\Ps_{\m_1\n_1|...|\,\m_n\n_n}$. Similar forms of the EoMs can be written for the Dirac conjugate spinor.

Finally, we have the following generalization of identity~\rf{x17}: \beq \de\cdot\mathcal{S}_{\m_1\dots\m_{n-1}}
=\tf{1}{2}\,\ds\,{\displaystyle{\not{\!\mathcal{S}}}}_{\m_1...\m_{n-1}}+\tf{n-1}{2}\,\de_{(\m_1}\mathcal{S}'
_{\m_2...\m_{n-1})}.\eeq{x27} This, when used in the divergence of Eq.~\rf{eomarb}, gives
\beq \de\cdot\mathcal{R}_{\m_1\dots\m_{n-1}}=-\tfrac{n-1}{2}\,\g_{(\m_1}\de
\cdot{\displaystyle{\not{\!\mathcal{S}\!\,}}}_{\m_2...\m_{n-1})}-\tfrac{(n-1)(n-2)}{4}\,\h_{(\m_1\m_2}\de
\cdot\mathcal{S}'_{\m_3...\m_{n-1})}.\eeq{Noether-n1}
Given Eqs.~\rf{noname4}--\rf{x20}, this can then be rewritten as
\beq \D\left(\de^{\m_n}\chi^*_{\m_1...\m_n}\right)=0,\eeq{Norther-n2} where
\beq \chi^*_{\m_1...\m_n}\equiv\ps^*_{\m_1...\m_n}-\tf{n-1}{2n+D-4}\,\g_{(\m_1}\pss^*_{\m_2...\m_{n-1})\m_n}
-\tf{(n-1)(n-2)}{2(2n+D-4)}\,\h_{(\m_1\m_2}\ps^{\prime*}_{\m_3...\m_{n-1})\m_n}.\eeq{Noether-n3}

%%%%%%%%%%%%%%%%%%%%%%%%%%%%%%%%%%%%%%%%%%%%%%%%%%%%%%
%%%%%%%%%%%%%%%%%%%%%%%%%%%%%%%%%%%%%%%%%%%%%%%%%%%%%%
\section{The Cohomology of $\G$}\label{sec:cohomology}
%%%%%%%%%%%%%%%%%%%%%%%%%%%%%%%%%%%%%%%%%%%%%%%%%%%%%%
%%%%%%%%%%%%%%%%%%%%%%%%%%%%%%%%%%%%%%%%%%%%%%%%%%%%%%

In this Appendix we clarify and prove some important facts about the cohomology of $\G$, used throughout the main
text. Let us recall that the action of $\G$ is defined by
%\
\bea &\G h_{\m\n}=2\de_{(\m} C_{\n)},&\label{Gammaaction1}\\&\G\ps_{\n_1...\n_n}=n\de_{(\n_1}\xi_{\n_2...\n_n)},
\qquad \G\bar\ps_{\n_1...\n_n}=-n\de_{(\n_1}\bar\xi_{\n_2...\n_n)}.&\eea{Gammaaction2}
The nontrivial elements in the cohomology of $\G$ are nothing but gauge-invariant objects that themselves are not
gauge variations of something else. Let us consider one by one all such elements, and also prove some useful relations
involving $\G$-exact terms.

%%%%%%%%%%%%%%%%%%%%%%%%%%%%%%%%%%%%%%%%%%%%%%%%
\subsection{Curvatures}\label{subsec:curvatures}
%%%%%%%%%%%%%%%%%%%%%%%%%%%%%%%%%%%%%%%%%%%%%%%%

The curvatures $\{R_{\m\n\r\s}, \Ps_{\m_1\n_1|...|\,\m_n\n_n}\}$ and their derivatives belong to $H(\G)$. That
the curvatures are $\G$-closed is easy to see. For the linearized Riemann tensor, $R_{\m\n\r\s}$, it follows from
the commutativity of partial derivatives as one takes the 2-curl of Eq.~\rf{Gammaaction1}:
\beq \G R_{\m\n}{}^{\r\s}=\G\left(4\de^{[\r}\de_{[\m}h_{\n]}{}^{\s]}\right)=4\de^{[\r}\de_{[\m}\de_{\n]}C^{\s]}
+4\de^{[\r}\de_{[\m}\de^{\s]}C_{\n]}=0.\eeq{g1}
One can also take a 1-curl of the first equation of~\rf{Gammaaction2} to obtain
\beq \G\,\uppsi^{(1)\m_1\n_1\Vert}{}_{\n_2...\n_n}=(n-1)\de_{(\n_2}\upxi^{(1)\m_1\n_1\|}{}_{\n_3...\n_n)},\eeq{g-1curl}
and similarly for the Dirac conjugate. Likewise, an $m$-curl of Eq.~\rf{Gammaaction2} gives, for $m\leq n$,
\beq \G\,\uppsi^{(m)\m_1\n_1|...|\m_m\n_m\Vert}{}_{\n_{m+1}...\n_n}=(n-m)\de_{(\n_{m+1}}
\upxi^{(m)\m_1\n_1|...|\m_m\n_m\Vert}{}_{\n_{m+2}...\n_{n)}}.\eeq{g-mcurl}
In particular, when $m=n$, we have the $\G$-variation of the curvature; it vanishes:
\beq \G\,\Ps^{\m_1\n_1|...|\m_n\n_n}=0.\eeq{g-ncurl}
Note that the $\G$-closure of the curvature holds without requiring any constraints on the fermionic ghost.
To see that the curvatures are not $\G$-exact, we simply notice that these are $pgh$-0 objects, whereas any
$\G$-exact piece must have $pgh>0$. Therefore, the curvatures are nontrivial elements in the cohomology of $\G$,
and so are their derivatives.

As we have already seen, only the highest curl ($n$-curl) of the field $\ps_{\n_1...\n_n}$ is $\G$-closed, while
no lower curls are. It is the commutativity of partial derivatives that plays a crucial role. Clearly, an arbitrary
derivative of the field will not be $\G$-closed in general. Yet, some particular linear combination of such
objects (or their $\g$-traces) can be $\G$-closed under the constrained ghost. The latter possibility is exhausted
precisely by the Fronsdal tensor and its derivatives, which will be discussed later.

%%%%%%%%%%%%%%%%%%%%%%%%%%%%%%%%%%%%%%%%%%%%%%%%
\subsection{Antifields}\label{subsec:antifields}
%%%%%%%%%%%%%%%%%%%%%%%%%%%%%%%%%%%%%%%%%%%%%%%%

The antifields  $\{h^{*\m\n}, C^{*\m}, \bar{\ps}^{*\m_1...\m_n}, \bar{\xi}^{*\m_1...\m_{n-1}}\}$ and their
derivatives belong to the cohomology of $\G$ as well. These objects are $\G$-closed simply because $\G$ does
not act on the antifields. On the other hand, having $pgh=0$, they cannot be $\G$-exact.

%%%%%%%%%%%%%%%%%%%%%%%%%%%%%%%%%%%%%%%%%%%%%%%%%%%%%%%%%%%
\subsection{Ghosts \& Ghost-Curls}\label{subsec:ghostcurls}
%%%%%%%%%%%%%%%%%%%%%%%%%%%%%%%%%%%%%%%%%%%%%%%%%%%%%%%%%%%

The undifferentiated ghosts $\{C_\m, \xi_{\m_1...\m_{n-1}}\}$ are $\G$-closed objects since $\G$ does not act on
them. Moreover, they cannot be $\G$-exact, thanks to Eqs.~\rf{Gammaaction1}--\rf{Gammaaction2}, which say
that any $\G$-exact term must contain at least one derivative of a ghost.

Any derivatives of the ghosts are also $\G$-closed. Some derivatives, though, will be $\G$-exact, i.e.,
trivial in the cohomology of $\G$. For example, any symmetrized derivatives of the bosonic ghost is
trivial: $\de_{(\m}C_{\n)}=\tf{1}{2}\G h_{\m\n}$, but its 1-curl is not. We have
\beq \de_\m C_\n=\de_{(\m}C_{\n)}+\de_{[\m}C_{\n]}=\tf{1}{2}\G h_{\m\n}+\tf{1}{2}\mathfrak{C}_{\m\n} .\eeq{g1.9}
By taking a curl of Eq.~\rf{Gammaaction1}, one however finds that any derivative of $\mathfrak{C}_{\m\n}$ is $\G$-exact:
\beq \de_\r\mathfrak{C}_{\m\n}=\G\mathfrak{h}_{\m\n\Vert\r}.\eeq{ghcurl}

Derivatives of the fermionic ghost are more interesting. In the simplest case of a spin-$\tf{5}{2}$ field, with $n=2$,
we see that
\beq \de_\m\xi_\n=\de_{(\m}\xi_{\n)}+\de_{[\m}\xi_{\n]}=\tf{1}{2}\G\ps_{\m\n}+\tf{1}{2}\upxi_{\m\n}.\eeq{g2}
The 1-curl $\upxi_{\m\n}$ is a nontrivial element in the cohomology of $\G$, but its $\g$-trace is not:
\beq \g^\a\upxi_{\a\b}=\ds\,\xi_\a=\G\pss_\a,\eeq{g}
thanks to the $\g$-tracelessness of the ghost. Again, a derivative of the 1-curl is trivial:
\beq \de_\r\upxi_{\m\n}=\G\,\uppsi_{\m\n\Vert\r},\qquad \de_\r\bar\upxi_{\m\n}=-\G\,\bar\uppsi_{\m\n\Vert\r},
\eeq{gxicurl}
which is obtained directly from Eq.~\rf{g-1curl} by setting $n=2$.
The $n=3$ counterpart of Eq.~\rf{g2} reads
\beq \de_\m\xi_{\n\r}=\de_{(\m}\xi_{\n\r)}+\tf{4}{3}\de_{[\m} \xi_{\n]\r}+\tf{2}{3}\de_{[\n}\xi_{\r]\m}=\tf{1}{3}
\G\ps_{\m\n\r}+\tf{2}{3}\upxi^{(1)}_{\m\n\Vert\r}+\tf{1}{3}\upxi^{(1)}_{\n\r\Vert\m}.\eeq{g3}
The generalization to arbitrary spin is straightforward. One obtains
\bea \de_\r\xi_{\n_1...\n_{n-1}}&=&\de_{(\r}\xi_{\n_1...\n_{n-1})}+2\left(1-\tf{1}{n}\right)\de_{[\r}
\xi_{\n_1]\n_2...\n_{n-1}}\nonumber\\&&+2\sum_{m=1}^{n-2}\left(1-\tf{m+1}{n}\right)\de_{[\n_m}
\xi_{\n_{m+1}]\r\,\n_1...\n_{m-1}\n_{m+2}...\n_{n-1}}\nonumber\\&=&\tf{1}{n}\,\G\ps_{\r\,\n_1...\m_{n-1}}
+\left(1-\tf{1}{n}\right)\upxi_{\r\n_1\|\n_2...\n_{n-1}}^{(1)}\nonumber\\&&+\sum_{m=1}^{n-2}\left(1-\tf{m+1}{n}\right)
\upxi_{\n_m\n_{m+1}\|\r\,\n_1...\n_{m-1}\n_{m+2}...\n_{n-1}}^{(1)}.\eea{g4}
We conclude that any first derivative of the fermionic ghost is a linear combination of 1-curls, up to $\G$-exact
terms. Therefore, it suffices to consider only 1-curls of the ghost in the cohomology of $\G$. More
generally, for $m$ derivatives, with $m\leq n-1$, one can consider only the $m$-curls in the cohomology of $\G$.
To see this, we can first take a curl of Eq.~(\ref{g4}) to convince ourselves that only 2-curls of the ghost
are nontrivial. Similarly, we can continue step by step to show that for any $m$-derivative combination of the fermionic ghost,
with $m\leq n-1$, it suffices to consider only $m$-curls thereof.

It is clear that the derivative of an $m$-curl, $\de_{\n_n}\upxi^{(m)}_{\m_1\n_1|...|\m_m\n_m\|\,\n_{m+1}...\n_{n-1}}$,
contains non-trivial $(m+1)$-curls. Only when symmetrized w.r.t. the indices $\{\n_{m+1}, ..., \n_n\}$, may this quantity
be $\G$-exact. This fact is nothing but a restatement of Eq.~(\ref{g-mcurl}) for $0\leq m\leq n-1$:
\beq \de_{(\n_n}\upxi^{(m)\m_1\n_1|...|\m_m\n_m\Vert}{}_{\n_{m+1}...\n_{n-1})}=\tf{1}{n-m}\,\G\,
\uppsi^{(m)\m_1\n_1|...|\m_m\n_m\Vert}{}_{\n_{m+1}...\n_n}.\eeq{exc}
Setting $m=n-1$, it follows immediately that a derivative of the highest ghost-curl is always $\G$-exact:
\beq \de_{\n_n}\upxi^{(n-1)}_{\m_1\n_1|...|\m_{n-1}\n_{n-1}}=\G\,\uppsi^{(n-1)}_{\m_1\n_1|...|\m_{n-1}\n_{n-1}
\Vert\n_n},\eeq{exc1}
which generalizes Eq.~(\ref{gxicurl}) for arbitrary spin.

However, the $\g$-trace of any $m$-curl, $\upxi^{(m)}_{\m_1\n_1|...|\m_m\n_m\|\n_{m+1}...\n_{n-1}}$,  is always
$\G$-exact. If the $\g$-matrix carries one of the unpaired indices $\{\n_{m+1},...,\n_{n-1}\}$, this quantity vanishes
since the ghost is $\g$-traceless. Otherwise, the same constraint gives rise to the following:
\beq \g^{\m_1}\upxi^{(m)}_{\m_1\n_1|...|\m_m\n_m\Vert\n_{m+1}...\n_{n-1}}=\ds\,\upxi^{(m-1)}_{\m_2\n_2|...|\m_m\n_m
\Vert\n_1\n_{m+1}...\n_{n-1}}.\eeq{g5}
But one can take a $\g$-trace of Eq.~(\ref{exc}) to see that the above quantity is actually $\G$-exact. Thus one
finds the arbitrary-spin generalization of Eq.~(\ref{g}):
\beq \g^{\m_1}\upxi^{(m)}_{\m_1\n_1|...|\m_m\n_m\Vert\n_{m+1}...\n_{n-1}}=\tf{1}{n-m}\,\G\!\displaystyle\not{\!\uppsi}
^{(m-1)}_{\m_1\n_1|...|\m_m\n_m\Vert\n_{m+1}...\n_{n-1}}.\eeq{g6}
So, one may exclude from the cohomology of $\G$ the $\g$-traces of the fermionic ghost-curls.

%%%%%%%%%%%%%%%%%%%%%%%%%%%%%%%%%%%%%%%%%%%%%%%%%%%
\subsection{Fronsdal Tensor}\label{subsec:fronsdal}
%%%%%%%%%%%%%%%%%%%%%%%%%%%%%%%%%%%%%%%%%%%%%%%%%%%

The Fronsdal tensor $\mathcal S_{\m_1...\m_n}$ and derivatives thereof also belong to the cohomology of $\G$.
From the definition, one finds that its $\G$ variation is given by
\bea \G\mathcal{S}_{\m_1...\m_n}&=&i\left[\ds\,\G \ps_{\m_1...\m_n}-n\de_{(\m_1}\G\pss_{\m_2...\m_n)}\right]\nonumber\\
&=& in\left[\ds\,\de_{(\m_1}\xi_{\m_2...\m_n)}-n\g^\r\de_{(\m_1}\de_{(\r}\xi_{\m_2...\m_n))}\right]\nonumber\\
&=&-in(n-1)\de_{(\m_1}\de_{(\m_2}\displaystyle{\not{\!\xi}}_{\m_3\dots\m_n))}.\eea{g7}
This quantity vanishes since the ghost is $\g$-traceless. $\mathcal S_{\m_1...\m_n}$, being a $pgh$-0 object, is
not $\G$-exact either. Therefore, the Fronsdal tensor and its derivatives belong to $H(\G)$.

In view of Eq.~\rf{eomarb2} and~\rf{eomarb4}, however, we see that the two highest curls of the Fronsdal tensor
boil down to objects already enlisted in Subsection~\ref{subsec:curvatures}, and therefore do not need separate
consideration. The aforementioned equations are generalizations of the Damour-Deser relations~\cite{DD,BB}.
Consequently, for the spin-$\tf{5}{2}$ case, it suffices to consider only symmetrized derivatives
of the Fronsdal tensor.

%%%%%%%%%%%%%%%%%%%%%%%%%%%%%%%%%%%%%%%%%%%%%%%%%%%%%%%%%%
%%%%%%%%%%%%%%%%%%%%%%%%%%%%%%%%%%%%%%%%%%%%%%%%%%%%%%%%%%
\section{Proof of Some Technical Steps}\label{sec:tedious}
%%%%%%%%%%%%%%%%%%%%%%%%%%%%%%%%%%%%%%%%%%%%%%%%%%%%%%%%%%
%%%%%%%%%%%%%%%%%%%%%%%%%%%%%%%%%%%%%%%%%%%%%%%%%%%%%%%%%%

Throughout the bulk of the paper, we have omitted the proof of some cumbersome technical steps for the sake of readability.
The detailed proof of those steps appear in this Appendix. We will use a number of $\g$-matrix identities, which can be
derived, for example, by using the Mathematica package called Gamma~\cite{Gamma}.

%%%%%%%%%%%%%%%%%%%%%%%%%%%%%%%%%%%%%%%%%%%%%%%%%%%%%%%%%%%%%%
\subsection{2-Derivative 2--5/2--5/2 Vertex}\label{subsec:C2d}
%%%%%%%%%%%%%%%%%%%%%%%%%%%%%%%%%%%%%%%%%%%%%%%%%%%%%%%%%%%%%%

In Eq.~\rf{1dera2}, the part of $a_2$ that contains the fermionic antighost is given by
\beq a_{2\tilde g}=-\tilde{g}\,\bar\xi^*_\r\g^{\a\b\r\m\n}\upxi_{\m\n}\mathfrak{C}_{\a\b}+\hc\,,\eeq{C.1}
which comes with five $\g$-matrices. But it can be cast into an equivalent form that contains just one, like that appearing
in Eq.~\rf{5halfa2}. To see this, let us first use the $\g$-matrix identity:
\beq \g^{\a\b\r\m\n}=\tf{1}{2}\left(\g^\a\g^{\b\r\m\n}+\g^{\b\r\m\n}\g^\a\right),\eeq{C.2}
and then another one for the antisymmetric product of four $\g$-matrices, namely
\beq \g^{\b\r\m\n}=-2\h^{\b\r|\m\n}+\g^{\b\r}\g^{\m\n}-4\g^{[\b}\h^{\r][\m}\g^{\n]}.\eeq{C.3}
The result is
\bea a_{2\tilde g}&=&-\tilde{g}\,\bar\xi^*_\r\g^\a\left(-\h^{\b\r|\m\n}+\tf{1}{2}\g^{\b\r}\g^{\m\n}-2\g^{[\b}\h^{\r]\m}
\g^{\n}\right)\upxi_{\m\n}\mathfrak{C}_{\a\b}\nonumber\\&&-\tilde{g}\,\bar\xi^*_\r\left(-\h^{\b\r|\m\n}+\tf{1}{2}\g^{\b\r}
\g^{\m\n}-2\g^{[\b}\h^{\r]\m}\g^{\n}\right)\g^\a\upxi_{\m\n}\mathfrak{C}_{\a\b}+\hc\,.\eea{C.4}
In both the first and the second lines on the right-hand side, the first term is of the desired form with a single $\g$-matrix,
while the second and third terms give rise to the $\G$-exact pieces $\g^{\n}\upxi_{\m\n}$ and  $\g^{\m\n}\upxi_{\m\n}$, either
directly or through the relations: $\g^{\m\n}\g^\a=\g^\a\g^{\m\n}-4\h^{\a[\m}\g^{\n]}$ and $\g^\n\g^\a=-\g^\a\g^\n
+2\h^{\n\a}$. Finally, on account of the $\g$-tracelessness of $\bar\xi^*_\r$, one obtains
\beq a_{2\tilde g}=-\tilde{g}\,\bar\xi^*_\r\left(-2\h^{\b\r|\m\n}\g^\a-2\g^{\b}\h^{\r\m}\h^{\n\a}\right)\upxi_{\m\n}
\mathfrak{C}_{\a\b}+\hc+\G\text{-exact},\eeq{C.5}
which is indeed equivalent to the $p=2$ piece presented in Eq.~\rf{5halfa2}, since more explicitly,
\beq a_{2\tilde g}=4\tilde{g}\,\bar\xi^{*\m}\g^\a\upxi^\b{}_\m\mathfrak{C}_{\a\b}+\hc+\G\text{-exact}.~\spadesuit\eeq{C.6}

Now we will fill up the gaps between Eqs.~\rf{beta1} and~\rf{beta5}. First we use the definition~\rf{N23} of $\chi^{*\m\n}$,
and Eqs.~\rf{nonmame2} to write
\beq \D\chi^*_{\r\s}=\mathcal{S}_{\r\s}-\tf{1}{2}\g_\s\displaystyle{\not{\!\mathcal{S}}}_{\r}-\tf{1}{2}\h_{\r\s}
\mathcal{S}'.\eeq{beta2}
One can take a curl of the above equation, and relate the 1-curl of the Fronsdal tensor to the $\g$-trace of the
curvature through Eq.~\rf{Fronsdal1curl}, which yields
\beq 2\D\de_{[\n}\chi^*_{\r]\s}=\tf{i}{2}\g_{\s\t\l}\Ps_{\n\r}{}^{\t\l}+\h_{\s[\n}\de_{\r]}\mathcal{S}'.\eeq{beta3}
When this expression is used in Eq.~\rf{beta1}, the $\mathcal{S}'$-terms vanish because of the Bianchi identity,
$\bar{\uppsi}_{[\a\b\Vert\n]}$, imposed by the antisymmetric 5-$\g$. The result is
\beq \b^\m_C~=~i\gt\,\bar\Ps_{\n\r|\t\l}\g^{\s\t\l}\g^{\m\n\r\a\b}\uppsi_{\a\b\Vert\s}~-~i\gt^*\bar
\uppsi_{\a\b\Vert\s}\g^{\m\n\r\a\b}\g^{\s\t\l}\Ps_{\n\r|\t\l}.\eeq{beta4}
Now one can take $\de_\t$ out of the curvature and use Leibniz rule to find a total derivative plus some terms
that can be identified as $-\b^\m_C$ only if $\gt$ is real. With this, one obtains
\beq \b^\m_C~=~2i\gt\,\de_\n\left(\bar\uppsi_{\r\s\Vert\l}\,\g^{\m\r\s\a\b,\,\n\l\g}\,\uppsi_{\a\b\Vert\g}\right).
\eeq{beta4.5}
Note that the quantity inside the parentheses can be made symmetric under $\m\leftrightarrow\n$ for free, thanks
to the Bianchi identities playing role when $\de_\n$ hits the 1-curls. This leads us to Eq.~\rf{beta5} under the stated
condition: $\tilde g$ is real.~$\spadesuit$

Next, we will derive Eq.~\rf{beta6.5} from Eq.~\rf{beta6}. We will simply show (dropping the quite similar proof for
hermitian conjugates) that
\beq 2ih_{\m\n}\,\G\,\bar\uppsi_{\r\s\Vert\l}\,\g^{\m\r\s\a\b,\,\n\l\g}\,\uppsi_{\a\b\Vert\g}-2\mathfrak{h}_{\m\n\Vert}
{}^\s\bar{\upxi}_{\a\b}\g^{\m\n\r\a\b}\D\chi^*_{\r\s}\doteq-i\bar\xi_\l R_{\m\n\r\s}\g^{\m\n\l\a\b,\,\t\r\s}\,
\uppsi_{\a\b\Vert\t}.\eeq{C.11}
Let us rewrite the left-hand side of Eq.~\rf{C.11} as
\beq \text{L.H.S.}=-2ih_{\m\n}\de_\l\bar{\upxi}_{\r\s}\,\g^{\m\r\s\a\b,\,\n\l\g}\,\uppsi_{\a\b\Vert\g}
-2\mathfrak{h}_{\m\n\Vert}{}^\s\bar{\upxi}_{\a\b}\g^{\m\n\r\a\b}\D\chi^*_{\r\s}.\eeq{C.12}
In the first term on the right-hand side, let us pull out the derivative $\de_\s$ off the ghost-curl and integrate by parts.
This is followed by another integration by parts w.r.t.~$\de_\l$. In the second term, on the other hand, we pull out the
derivative $\de_\b$ off the ghost-curl to integrate by parts. In these steps, we exploit the antisymmetry of the products
of $\g$-matrices, which kills some terms by enforcing the second Bianchi identities given in Eqs.~\rf{x1} and~\rf{x11}.
Then, we are left with
\beq \text{L.H.S.}\doteq i\bar{\xi}_\r\left(\mathfrak{h}_{\s\n\Vert\m}\,\g^{\s\n\r\a\b,\,\m\l\g}\,\Ps_{\a\b|\l\g}+
R_{\s\n\l\m}\,\g^{\s\n\r\a\b,\,\m\l\g}\,\uppsi_{\a\b\Vert\g}\right)-4\bar{\xi}_\a\mathfrak{h}_{\m\n\Vert}
{}^\s\g^{\m\n\r\a\b}\D\de_{[\b}\chi^*_{\r]\s}.\eeq{C.13}
In the last term on the right-hand side above, one can again use Eq.~\rf{beta3} and then drop the $\mathcal{S}'$-terms
on account of the Bianchi identities. The result is
\beq \text{L.H.S.}\doteq i\bar{\xi}_\r\left(\mathfrak{h}_{\s\n\Vert\m}\,\g^{\s\n\r\a\b,\,\m\l\g}\,\Ps_{\a\b|\l\g}+
R_{\s\n\l\m}\,\g^{\s\n\r\a\b,\,\m\l\g}\,\uppsi_{\a\b\Vert\g}\right)-i\bar{\xi}_\a\mathfrak{h}_{\m\n\Vert\s}
\g^{\m\n\r\a\b}\g^{\s\t\l}\Ps_{\b\r|\t\l}.\eeq{C.14}
Now, in the last term on the right-hand side, the matrices $\g^{\m\n\r\a\b}$ and $\g^{\s\t\l}$ actually commute. This can
be seen by first writing $\g^{\s\t\l}=\tf{1}{2}\left(\g^\s\g^\t\g^\l-\g^\l\g^\t\g^\s\right)$, and then noticing that
any of these $\g$-matrices commutes past $\g^{\m\n\r\a\b}$, on account of the identity:
\beq \g^{\m\n\r\a\b}\g^\s=\g^\s\g^{\m\n\r\a\b}-2\g^{\s\m\n\r\a\b},\eeq{C.15} and similar ones for $\g^\t$ and $\g^\l$,
and the fact that the antisymmetric products of six $\g$-matrices are always eliminated by the Bianchi identities.
This enables us to rewrite the last term on the right-hand side of Eq.~\rf{C.14} as the first one, but with an opposite
sign, so that these terms actually cancel each other. Therefore, we are left only with
\beq \text{L.H.S.}\doteq-i\bar\xi_\l R_{\m\n\r\s}\g^{\m\n\l\a\b,\,\t\r\s}\,\uppsi_{\a\b\Vert\t}.\eeq{C.16}
This is precisely the right-hand side of Eq.~\rf{C.11}, which, therefore, is proved.~$\spadesuit$

Now we will show how Eq.~\rf{beta7} follows from Eq.~\rf{beta6.5}. In other words, we will prove
\beq -i\gt\,\bar\xi_\l R_{\m\n\r\s}\g^{\m\n\l\a\b,\,\t\r\s}\,\uppsi_{\a\b\Vert\t}+\hc\doteq-8i\gt\,\G\left(\bar\ps_{\m\a}
R^{+\m\n\a\b}\ps_{\n\b}+\tf{1}{2}\bar{\pss}_\m\Rs^{\m\n}\pss_\n\right)-\D\mathfrak{a},\eeq{C.17}
for some $\D\mathfrak{a}$ to be determined. First, let us write down a $\g$-matrix identity:
\beq \g^{\m\n\l\a\b,}{}_{\t\r\s}=-60\,\d^{[\m\n\l}_{\;\t\r\s}\,\g^{\a\b]}\,+\,15\,\d^{[\m}_{[\t}\,\g_{\r\s]}
{}^{\n\l\a\b]}.\eeq{C.18}
Using this identity, one can rewrite the left-hand side of Eq.~\rf{C.17} as
\beq \text{L.H.S.}=60i\gt\,\bar\xi_\l\,\d^{[\m\n\l}_{\;\r\s\t}\,\g^{\a\b]}R_{\m\n}{}^{\r\s}\,\uppsi_{\a\b\Vert}
{}^{\t}+\hc\,,\eeq{C.19}
where the potential terms with six $\g$-matrices have all been eliminated by the Bianchi identities. Whenever the
index $\l$ is on a $\g$-matrix, we will get rid of $\g$-matrices altogether by using the identity:
$\g^{\l\a}=\g^\l\g^\a-\h^{\l\a}$, and the $\g$-tracelessness of the fermionic ghost. Otherwise, if just one of the
indices $\a$ and $\b$ appears on a $\g$-matrix, we will use the same identity to obtain a single $\g$-trace of
$\uppsi_{\a\b\Vert}{}^{\t}$. These steps leave us with
\bea \text{L.H.S.}&=&12i\gt\,\bar\xi_\l R_{\m\n}{}^{\r\s}\left(\d^{\a\b\m}_{\r\s\t}\,\h^{\n\l}+\d^{\m\n\a}_{\r\s\t}\,
\h^{\b\l}+2\d^{\l\m\a}_{\r\s\t}\,\h^{\n\b}+\tf{1}{2}\d^{\l\m\n}_{\r\s\t}\,\g^{\a\b}\right)\,\uppsi_{\a\b\Vert}{}^{\t}
+\hc\nonumber\\&&-24i\gt\,\bar\xi_\l R_{\m\n}{}^{\r\s}\,\d^{\l\m\a}_{\r\s\t}\,\g^\n\g^\b\,\uppsi_{\a\b\Vert}{}^{\t}
+6i\gt\,\bar\xi_\l\,\d^{\l\a\b}_{\r\s\t}\,\Rs^{\r\s}\uppsi_{\a\b\Vert}{}^{\t}+\hc\,.\eea{C.20}

It is rather easy to see that the entire first line reduces, up to total derivatives, to a $\G$-exact piece modulo
$\D$-exact terms. Although more difficult to see, the same is true for the second line as well. Let us call the first and
the second lines on the right-hand side of Eq.~\rf{C.20} respectively as 1st Line and 2nd Line. In 1st Line we can
use the relation~\rf{x14}, and carry out an explicit computation to write down
\bea \text{1st Line}&=&-8i\gt\,\bar\xi_\l\left[\left(R^{\l\n\a\b}+R^{\a\n\l\b}\right)\de_\a\ps_{\n\b}+2R^{\l\b}
\uppsi_{\a\b\Vert}{}^\a-R^{\a\b}\uppsi^\l{}_{\a\Vert\b}+\tf{1}{2}R\uppsi^{\l\a\Vert}{}_\a\right]\nonumber\\
&&+6i\gt\,\bar\xi_\l\left[iR_{\m\n}{}^{[\l\m}\displaystyle{\not{\!\mathcal{S}}}^{\n]}-\de^{[\l}\left(R_{\m\n}{}^{\m\n]}
\ps'\right)\right]+\hc\,.\eea{C.21}
One can integrate by parts w.r.t.~$\de_\a$ in the terms containing the Riemann tensor, and thereby extract a $\G$-exact
piece. The result is
\bea \text{1st Line}&\doteq&-8i\gt\,\G\left(\bar\ps_{\m\a}R^{\m\n\a\b}\ps_{\n\b}\right)\nonumber\\
&&+16i\gt\left[\bar\xi_\l\left(\de_\a
R^{\l\n\a\b}\ps_{\n\b}-R^{\l\b}\uppsi_{\a\b\Vert}{}^\a+\tf{1}{2}R^{\a\b}\uppsi^\l{}_{\a\Vert\b}-\tf{1}{4}R\uppsi^{\l\a
\Vert}{}_\a\right)-\hc\right]\nonumber\\&&+2i\gt\left[3i\bar\xi_\l R_{\m\n}{}^{[\l\m}\displaystyle{\not{\!\mathcal{S}}}
^{\n]}-\bar\xi_\l\de^\l\left(R\ps'\right)+2\bar\xi_\l\de_\m\left(R^{\m\l}\ps'\right)-\hc\right],\eea{C.22}
which is manifestly of the form $\G$-exact plus $\D$-exact.

Similarly, in the first term of 2nd Line, one can use Eq.~\rf{x13}. The result is
\beq \text{2nd Line}=24i\gt\,\bar\xi_\l\g^\m \left[iR_{\m\n}{}^{[\a\l}\mathcal{S}_\a{}^{\n]}-\de^{[\l}\left(R_{\m\n}
{}^{\n\a]}\pss_\a\right)\right]+12i\gt\,\bar\xi_\l\Rs^{[\a\b}\de_\a\ps_\b{}^{\l]}+\hc\,.\eeq{C.23}
The second term in the brackets contains manifestly $\D$-exact pieces, which we separate:
\bea \text{2nd Line}&=&8i\gt\,\bar\xi_\l\g^\m \left[3iR_{\m\n}{}^{[\a\l}\mathcal{S}_\a{}^{\n]}+\de^\l\left(R_{\m\a}
\pss^\a\right)-\de^\a\left(R_\m{}^\l\pss_\a\right)\right]+\hc\nonumber\\&&+4i\gt\,\bar\xi_\l\left[3\Rs^{[\a\b}\de_\a
\ps_\b{}^{\l]}-2\g^\m\de^\n\left(R_{\m\n}{}^{\a\l}\pss_\a\right)\right]+\hc\,.\eea{C.24}
The first line on the right-hand side is now manifestly $\D$-exact, whereas the second line can be written as $\G$-exact
plus $\D$-exact modulo total derivatives, which we will now show.

To this end, we will first compute the $\G$ variation of the following quantity:
\beq Z\equiv-4i\gt\left(\bar\ps_{\l\a}\g^{\l\n\r\s}R_{\r\s}{}^{\a\b}\ps_{\n\b}+\bar{\pss}_\m\Rs^{\m\n}\pss_\n\right).
\eeq{C.25}
The $\G$ variation gives derivatives of the ghost, but integrations by parts will yield
\bea \G Z&\doteq&-2i\gt\,\bar\xi_\l\left[\g^{\l\n\r\s}R_{\r\s}{}^{\a\b}\uppsi_{\a\b\Vert\n}+\g^{\a\n\r\s}R_{\r\s}
{}^{\l\b}\uppsi_{\a\n\Vert\b}\right]\nonumber\\&&~~~~~~~~~~~~~~~~-4i\gt\,\bar\xi_\l\g^{\l\n\r\s}\de_\a R_{\r\s}
{}^{\a\b}\ps_{\n\b}-4i\gt\bar\xi_\l\ds\left(\Rs^{\l\a}\ps_\a\right)+\hc\,.\eea{C.26}
Let us use the $\g$-matrix identity: $\g^{\l\n\r\s}=2\h^{\l\n|\r\s}+\tf{1}{2}\left(\g^{\l\n}\g^{\r\s}+
\g^{\r\s}\g^{\l\n}\right)$ for the first term in the brackets, and $\g^{\a\n\r\s}=-2\h^{\a\n|\r\s}+\tf{1}{2}
\left(\g^\a\g^{\r\s}\g^\n-\g^\n\g^{\r\s}\g^\a\right)$ for the second one. Furthermore, we break $\g^{\l\n}$
to obtain the $\g$-trace of either the ghost (which is zero) or the fermion 1-curl (for which we use Eq.~\rf{x13}).
The result is
\bea \G Z&\doteq&2i\gt\,\bar\xi_\l\left[\left(\g^\b\Rs^{\l\a}-\Rs^{\a\b}\g^\l\right)\de_\a\pss_\b-2\de_\b\left(
\g^\b\Rs^{\l\a}\pss_\a\right)\right]+4i\gt\,\bar\xi_\l\Rs^{\a\b}\de_\a\ps_\b{}^\l\nonumber\\&&~~~~~~~~~~~~~~~~
-2i\gt\,\bar\xi_\l\left(i\g^\a\Rs^{\l\b}\mathcal S_{\a\b}+2\g^{\l\n\r\s}\de_\a R_{\r\s}{}^{\a\b}\ps_{\n\b}\right)
+\hc\,.\eea{C.27}
In the first line above, for all three quantities inside the brackets, we commute the $\g$-matrix past the double
$\g$-trace of the Riemann tensor. This leaves us with
\bea \G Z&\doteq&4i\gt\,\bar\xi_\l\left[3\Rs^{[\a\b}\de_\a\ps_\b{}^{\l]}-2\g^\m\de^\n\left(R_{\m\n}{}^{\a\l}\pss_\a\right)
\right]+2\gt\,\bar\xi_\l\left(\g^\a\Rs^{\l\b}\mathcal S_{\a\b}+\Rs^{\l\a}\displaystyle{\not{\!\mathcal S}}_\a\right)
\nonumber\\&&-4i\gt\,\bar\xi_\l\left(\g^{\l\n\r\s}\de_\a R_{\r\s}{}^{\a\b}\ps_{\n\b}+2\g_\s\de_\a R^{\l\b\a\s}
\pss_\b+\Rs^{\l\a}\overset\leftarrow\ds\,\ps_\a\right)+\hc\,.\eea{C.28}
Combining all the results, i.e., Eqs.~\rf{C.20},~\rf{C.22},~\rf{C.24} and~\rf{C.28}, we finally arrive at
Eq.~\rf{C.17}, where $\D\mathfrak a$ is given by
\bea \D\mathfrak{a}&=&-16i\gt\,\bar\xi_\l\left[\de_\a R^{\l\n\a\b}\ps_{\n\b}-R^{\l\b}\uppsi_{\a\b\Vert}{}^\a
+\tf{1}{2}R^{\a\b}\uppsi^\l{}_{\a\Vert\b}-\tf{1}{4}R\uppsi^{\l\a\Vert}{}_\a\right]\nonumber\\&&-8i\gt\,\bar\xi_\l
\left[\de^\l\left(\g^\m R_{\m\a}\pss^\a\right)-\de^\a\left(\g^\m R_\m{}^\l\pss_\a\right)+\tf{1}{2}\de_\m
\left(R^{\m\l}\ps'\right)-\tf{1}{4}\de^\l\left(R\ps'\right)\right]\nonumber\\&&-4i\gt\,\bar\xi_\l\left[\g^{\l\n\r\s}
\de_\a R_{\r\s}{}^{\a\b}\ps_{\n\b}+2\g_\s\de_\a R^{\l\b\a\s}\pss_\b+\Rs^{\l\a}\overset\leftarrow\ds\,\ps_\a\right]
\nonumber\\&&+2\gt\,\bar\xi_\l\left[3R_{\m\n}{}^{[\l\m}\displaystyle{\not{\!\mathcal{S}}}^{\n]}+12\g^\m R_{\m\n}
{}^{[\a\l}\mathcal{S}_\a{}^{\n]}+\g^\a\Rs^{\l\b}\mathcal S_{\a\b}+\Rs^{\l\a}\displaystyle{\not{\!\mathcal S}}
_\a\right]+\hc\eea{Dfraka}
This completes our proof.~$\spadesuit$

Having found $\D\mathfrak a$, we will now see how this quantity may be related to $\D a_{1g}$, given by
Eq.~\rf{Da1g}. This will lead us to the desired relation~\rf{beta8}. Note from Eq.~\rf{Da1g} that the
graviton EoMs in $\D a_{1g}$ appear only through the Einstein tensor $G^{\m\n}$. Therefore, we will
rewrite all the $\D$-exact terms in the first, second and third lines on the right-hand side of
Eq.~\rf{Dfraka} in terms of the Einstein tensor, by making use of the relations~\rf{grav1}--\rf{x3}.
For he antisymmetric 4-$\g$, we use the identity~\rf{C.3} in order to kill some terms that give the
$\g$-trace of $\bar\xi_\l$. We find that all the terms proportional to the trace of the Einstein
tensor (Ricci scalar) combine into $\G$-exact pieces. After some simplifications, the result is
\bea \D\mathfrak{a}&\doteq&8i\gt\,\bar\xi_\l\left[2G^{\m\n}\de^\l\ps_{\m\n}-3\de_\m\left(G^{\m\n}\ps_\n
{}^\l\right)+\de^\n\left(G^{\l\m}\ps_{\m\n}\right)\right]\nonumber\\&&+4\gt\,\bar\xi_\l\left[\tf{3}{2}R_{\m\n}
{}^{[\l\m}\displaystyle{\not{\!\mathcal{S}}}^{\n]}+6\g^\m R_{\m\n}{}^{[\a\l}\mathcal{S}_\a{}^{\n]}+\g^{\a\r\s}
R_{\r\s}{}^{\l\b}\mathcal S_{\a\b}+2\displaystyle{\not{\!G}}_\a\mathcal S^{\a\l}-G^{\b\l}\displaystyle{
\not{\!\mathcal S}}_\b\right]\nonumber\\&&+4i\gt\,\G\left[\bar\pss_\m\displaystyle{\not{\!G}}_\n\ps^{\m\n}
+\bar\ps_{\m\n}\displaystyle{\not{\!G}}^\m\pss^\n-\bar\pss_\m G^{\m\n}\pss_\n+2\bar\ps_{\m\a}G^{\m\n}\ps_\n
{}^\a\right]\nonumber\\&&-6i\gt\,\G\left[\bar\ps'G^{\m\n}\ps_{\m\n}+\bar\ps_{\m\n}G^{\m\n}\ps'-\tf{1}{3}
\bar\ps_{\m\n}R\ps^{\m\n}+\tf{1}{2}\bar\ps'R\ps'\right]+\hc\,.\eea{C.30}
The entire first line on the right-hand side plus its hermitian conjugate is easily identified, up to an
overall factor, as $\D a_{1g}$.  All the remaining terms, on the other hand, are $\D$ variations of $\G$-closed
quantities, and can be identified as $\D\tilde a_1$. Explicitly,
\bea \tilde a_1&=&4\gt\,\bar\xi_\l\left(\tf{3}{2}R_{\m\n}{}^{[\l\m}\displaystyle{\not{\!\vf}}^{\,*\n]}
+6\g^\m R_{\m\n}{}^{[\a\l}\vf^*_\a{}^{\n]}+\g^{\a\r\s}R_{\r\s}{}^{\l\b}\vf_{\a\b}^*+2\displaystyle{\not{\!G}}
_\a\vf^{*\a\l}-G^{\b\l}\displaystyle{\not{\!\vf}}^*_\b\right)\nonumber\\&&+8i\gt\,\G\left(\bar\pss_\m
\displaystyle{\not{\!h}}^*_\n\ps^{\m\n}-\tf{1}{2}\bar\pss_\m h^{*\m\n}\pss_\n+\bar\ps_{\m\a}h^{*\m\n}\ps_\n
{}^\a-\tf{3}{2}\bar\ps'h^{*\m\n}\ps_{\m\n}\right)\nonumber\\&&-\left(\tf{4}{D-2}\right)i\gt\,\G\left(
\bar\ps_{\m\n}h^{*\prime}\ps^{\m\n}-\tf{3}{2}\bar\ps' h^{*\prime} \ps'\right)+\hc\,.\eea{2deramb}
Thus we have proved the relation~\rf{beta8}, where the ambiguity is given by the above expression.~$\spadesuit$.

%%%%%%%%%%%%%%%%%%%%%%%%%%%%%%%%%%%%%%%%%%%%%%%%%%%%%%%%%%%%%%
\subsection{3-Derivative 2--5/2--5/2 Vertex}\label{subsec:C3d}
%%%%%%%%%%%%%%%%%%%%%%%%%%%%%%%%%%%%%%%%%%%%%%%%%%%%%%%%%%%%%%

First, we will show that the $a_2$ presented in Eq.~\rf{3der1} is equivalent to that appearing in the third line of
Eq.~\rf{5halfa2-C}. Given the identities~\rf{C.2} and~\rf{C.3}, we rewrite Eq.~\rf{3der1} as
\bea a_2&=&-ig\,C^*_\l\bar{\upxi}_{\m\n}\g^\l\left(-\h^{\m\n|\a\b}+\tf{1}{2}\g^{\m\n}\g^{\a\b}-2\g^{\m}\h^{\n\a}
\g^{\b}\right)\upxi_{\a\b}\nonumber\\&&-ig\,C^*_\l\bar{\upxi}_{\m\n}\left(-\h^{\m\n|\a\b}+\tf{1}{2}\g^{\m\n}
\g^{\a\b}-2\g^{\m}\h^{\n\a}\g^{\b}\right)\g^\l\upxi_{\a\b}.\eea{D.1}
It is clear that only the first terms in both the lines on the right-hand side are nontrivial, since the $\g$-trace
of the ghost-curl $\upxi_{\a\b}$ is $\G$-exact. This leaves us with
\beq a_2=2ig\,C^*_\l\,\bar{\upxi}_{\m\n}\g^\l\upxi^{\m\n}+\G\text{-exact},\eeq{D.2}
thereby proving the claimed equivalence.~$\spadesuit$

Now we will prove the statements that follow Eq.~\rf{3der5}. Let us take the first term on the right-hand side of
Eq.~\rf{3der5}, and use the identity~\rf{C.3} to rewrite it as
\beq igR_{\m\n\r\s}\bar{\xi}_{\l}\,\g^{\l\m\n\a\b}\,\Ps_{\a\b|}{}^{\r\s}=igR_{\m\n\r\s}\bar{\xi}_{\l}\,\g^{\l\m\n\a\b}\,
\left(-\tf{1}{2}\g^{\r\s\g\d}+\tf{1}{2}\g^{\r\s}\g^{\g\d}-2\g^{[\r}\h^{\s][\g}\g^{\d]}\right)\Ps_{\a\b|\g\d}.\eeq{D.3}
The first term on the right-hand side plus its hermitian conjugate is $\G$-exact modulo $d$, while the remaining terms are
$\D$-exact. To see this, let us massage these terms. We have
\beq \text{First Term}+\hc=-\tf{i}{2}gR_{\m\n\r\s}\left(\bar{\xi}_\l\g^{\l\m\n\a\b,\,\r\s\g\d}\,\Ps_{\a\b|\g\d}
-\bar{\Ps}_{\a\b|\g\d}\,\g^{\l\m\n\a\b,\,\r\s\g\d}\xi_\l\right),\eeq{D.4}
by virtue of the fact that the antisymmetric products of 5-$\g$ and 4-$\g$ commute for exactly the same reason as presented
in between Eqs.~\rf{C.14} and~\rf{C.16}. Now we can pull $\de_\m$ off the Riemann tensor to integrate by parts. Because of
the Bianchi identities, we get
\beq \text{First Term}+\hc\doteq-\tf{i}{2}g\,\mathfrak{h}_{\r\s\Vert\l}\left(\bar{\upxi}_{\m\n}\g^{\l\m\n\a\b,\,\r\s\g\d}\,
\Ps_{\a\b|\g\d}-\bar{\Ps}_{\a\b|\g\d}\,\g^{\l\m\n\a\b,\,\r\s\g\d}\upxi_{\m\n}\right),\eeq{D.5}
Finally, we pull $\de_\g$ off the spin-$\tf{5}{2}$ curvature and integrate by parts to obtain a derivative $\de_\g\upxi_{\m\n}$
of the ghost-curl, which is $\G$-exact. Thus we end up having
\beq \text{First Term}+\hc\doteq-ig\,\G\left(\mathfrak{h}^{\r\s\Vert\l}\bar{\uppsi}_{\m\n\Vert\g}\,\g^{\l\m\n\a\b,\,\r\s\g\d}\,
\uppsi_{\a\b\Vert\d}\right).\eeq{D.6}

On the other hand, it is manifest that the second and third terms appearing on the right-hand side of Eq.~\rf{3der5} are
$\D$-exact quantities. Moreover, they are $\D$ variations of some $\G$-closed objects. The following choice of
the ambiguity will eliminate these terms:
\beq \D\tilde a_1=-igR_{\m\n\r\s}\bar{\xi}_{\l}\,\g^{\l\m\n\a\b}\,\left(\tf{1}{2}\g^{\r\s}\g^{\g\d}-2\g^{[\r}
\h^{\s][\g}\g^{\d]}\right)\Ps_{\a\b|\g\d}+\hc\,.\eeq{D.7}
This choice is tantamount to
\beq \tilde a_1=-gR_{\m\n\r\s}\,\bar{\xi}_{\l}\left(4\g^{\l\m\n\a\b,\,\r}\,\de_{[\a}\vf^*_{\b]}{}^{\s}+\tf{1}{D}\,
\g^{\l\m\n\a\b,\,\r\s}\displaystyle{\not{\!\uppsi}}^*_{\a\b}\right)+\hc\,,\eeq{3der6}
and with this we arrive at Eq.~\rf{3der7}.~$\spadesuit$

\end{appendix}

\end{document}